\documentclass[aps,physrev,preprint,groupedaddress]{revtex4-2}

\usepackage{graphicx}%
\usepackage{multirow}%
\usepackage{amsmath,amssymb,amsfonts}%
\usepackage{amsthm}%
\usepackage{mathrsfs}%
\usepackage[title]{appendix}%
\usepackage{xcolor}%
\usepackage{textcomp}%
\usepackage{booktabs}%
\usepackage{algorithm}%
\usepackage{algorithmicx}%
\usepackage{algpseudocode}%
\usepackage{listings}%
\usepackage{subcaption}
\usepackage{tikz}
\usetikzlibrary{decorations.pathmorphing}
\usepackage{comment}

\begin{document}



\title{AdS black holes in two-dimensional dilaton gravity and holography}


\author{Uriel Noriega-Cornelio}
\email[]{uriel.noriegacor1@alumno.buap.mx}
\affiliation{Facultad de Ciencias Físico Matemáticas, Benemérita Universidad Autónoma de Puebla, Ciudad Universitaria, Puebla, CP 72570, Puebla, México}

\author{Alfredo Herrera-Aguilar}
\email[]{aherrera@ifuap.buap.mx}
\affiliation{Instituto de Física, Benemérita Universidad Autónoma de Puebla, Edificio IF-1, Ciudad Universitaria, Puebla, CP 72570, Puebla, México}

\author{Cupatitzio Ramírez-Romero}
\email[]{cramirez@fcfm.buap.mx}
\affiliation{Facultad de Ciencias Físico Matemáticas, Benemérita Universidad Autónoma de Puebla, Edificio FM-2, Ciudad Universitaria, Puebla, CP 72570, Puebla, México}


\date{\today}

\begin{abstract}
In this paper, we present two novel analytic AdS black hole solutions in a two-dimensional dilaton gravity theory with two scalar fields non-minimally coupled to gravity. Our solutions contain two arbitrary integration constants in the blackening factor $f(r)$, allowing for an extremal configuration. Solution I reproduces  a previously reported AdS black hole when one of the integration constants in $f(r)$ vanishes. For our black hole configurations, the scalar curvature is constant and negative, corresponding to the $AdS_2$ spacetime. In order to elucidate their black hole nature, we explore the causal structure of these solutions with the aid of suitable Kruskal-like coordinates and Penrose diagrams. By employing the Hamilton-Jacobi method, we construct a boundary counter-term that renders a renormalized action with a vanishing variation. We use this finite action for the partition function in the semi-classical approximation. We establish a consistent Thermodynamics, verified by the first law, for our black hole solutions, including the extremal case. Finally, we perform a holographic analysis of the effective theory at the boundary of the black hole solution I. This theory is characterized by a Schwarzian action supplemented by a black hole mass term determined by the two integration constants in $f(r)$. We also examine the holographic implications of the boundary counter-term.
\end{abstract}


\maketitle

\section{\label{sec:intro}Introduction}

The gauge/gravity correspondence \cite{Maldacena}, based on the relationship between a gravitational background and a quantum field theory living at the boundary, has shown to be a valuable tool for studying strongly coupled field theories. 
         
        In the gauge/gravity duality, the nongravitational system in thermal equilibrium at temperature $T$  is in direct correspondence with a black hole with the Hawking temperature $T$. An AdS black hole in $D+1$ dimensions
        is described by 
        \begin{equation} \label{}
        ds=l^2 \left(-r^{2}\;f(r)\;dt^{2}+\frac{dr^2}{r^2\;f(r)}+r^{2} \;dx_i dx^i\right),
        \end{equation}
        where $l$ stands for the $AdS$ radius, $r$ is the holographic or extra coordinate, $x_i$ labels the coordinates in the spatial sector with $i=1,2,...D-1$, and the function $f(r)$, known as the blackening factor, must asymptotically approach unity in order to recover the $AdS$ background at infinity. 
        

    In this paper we present two families of $AdS_2$ black hole analytical solutions  within the framework of a two-dimensional dilaton gravity theory with two scalar fields non-minimally coupled to gravity and a metric given by
\begin{equation} \label{ansatz}
    ds=l^2 \left(-r^{2}\;f(r)\;dt^{2}+\frac{dr^2}{r^2\;f(r)}\right).
\end{equation}
Two-dimensional black holes have been studied extensively in the literature, as they serve as models for testing ideas about the physics and Thermodynamics of black holes, with the intention to give insight into quantum gravity in higher dimensions. The novel $AdS_2$ black hole solutions coupled to scalar fields presented here pursue the aim of contributing in this direction by incorporating in the blackening factor an extra constant of integration into play and by serving as toy configurations that could be straightforwardly generalized to higher dimensions.

Two-dimensional anti-de Sitter, $AdS_2$, is a maximally symmetric spacetime invariant under the $SO(2, 1)$ group. This invariance is relevant in the AdS/CFT duality because it corresponds to the conformal invariance of the theory living on the $AdS_2$ boundary.

$AdS_2$ holography has been extensively considered since the early days of the AdS/CFT correspondence, see for instance \cite{Strominger1, Cadoni, Cadoni2, Michelson, Spradlin, Navarro, Azeyanagi, Chamon, Hartman, Leston}. In pure $AdS_2$ gravity, any finite-energy excitation above the vacuum solution induces a strong backreaction on the spacetime, leaving the ground states as the only consistent configurations \cite{Michelson}. However,  in \cite{Almheiri} it was shown that the gravitational backreaction can be regulated by considering a non-trivial dilaton field in the Jackiw-Teitelboim (J-T) model \cite{Teitelboim, Jackiw}, which allows for finite-energy states.

In J-T gravity, the nearly $AdS_2$ geometry leads to an effective boundary theory described by a Schwarzian action that is invariant under the $SL(2,R)$ symmetry \cite{Almheiri, Stanford, Engelsoy}. It has been shown that this very action describes models of quantum mechanical Majorana fermions, as is the case of the Sachdev-Ye-Kitaev (SYK) theory \cite{Sachdev, Kitaev}. The duality between SYK and $AdS_2$ gravity was first proposed in \cite{Sachdev2}; further studies include \cite{Sachdev3, Rosenhaus, Suzuki,Stanford3}. Taking into account the two constants of integration in the blackening factor, we analyze the effective theory living at the boundary of our black hole configurations, the corresponding solutions and their thermodynamical properties.

 Furthermore, the $AdS_2$ holography has provided an unparalleled toolkit for studying quantum chaos, see, e.g., \cite{Shenker2, Shenker3, Jensen, Stanford, Polchinski, Engelsoy}. In the boundary theory, out-of-time-ordered four-point functions exhibit exponential growth with a Lyapunov-type behavior that saturates the chaos bound established in \cite{Shenker}.

 On the other hand, dilaton gravity models with one or multiple scalar fields have been used to study holographically non-conformal field theories with an underlying generalized conformal structure \cite{Marikasyk}. These theories are scale invariant, provided that their couplings also scale. The structure of these models is captured by Ward identities, relating the stress-energy tensor and scalar operators, which imply restrictions to the correlation functions of the scalar operators. The SYK theory has an associated generalized conformal structure, studied holographically by two-dimensional dilaton gravity models with one or multiple scalar fields. It has been shown that the Ward identities governing this structure are obtained by means of the dual two-dimensional dilaton gravity theories. 

As exemplified above, two-dimensional dilaton gravity models are frequently found in the literature as a result of dimensional reduction of systems defined in higher dimensions. The most known example is the extremal Reissner-Nordstr\"{o}m (RN) black hole solution of four-dimensional Einstein-Maxwell gravity (see, for instance, \cite{Stanford}). In this example, near $AdS_2$ spacetime arises as the near-horizon geometry of the near-extremal RN black hole. For our two-dimensional black hole solutions, the extremality condition arises as a relation  between the constants of integration in the metric, even in the absence of an electric charge. Interestingly, in this extremal scenario, pure $AdS_2$ spacetime emerges in the whole manifold, not only near the horizon.

    
    Thus, with the aim of studying the global spacetime structure of our black hole solutions, we employ an Eddington-Finkelstein type transformation of coordinates to show the causal structure of spacetime through the behavior of the light cones. Then, we construct a set of Kruskal patches and corresponding diagrams, adequate to explore the regions containing the outer and inner horizons. We depict interesting properties of these spacetimes making use of the appropriate Penrose diagram.
    
    
    In particular, it is possible to deduce the thermodynamic quantities for black holes in two dimensions using the Euclidean path integral approximation for the partition function \cite{Gibbons}
\begin{equation} \label{aproximation partition function}
\mathcal{Z} \sim \exp \left(-\frac{1}{\hbar} I_E\right),
\end{equation}
where $I_E$ represents the Euclidean action evaluated in the classical solutions of  the field equations and $\hbar$ is the Planck constant. In order to use the saddle point approximation in (\ref{aproximation partition function}), it is necessary to have an action that is finite on-shell and whose variation $\delta I$ vanishes. In general, these conditions are not necessarily met by an action. As in higher dimensions, the on-shell action might diverge. This issue is commonly solved by the method of background subtraction \cite{Gibbons}, \cite{Liebl}, which has been applied, for example, to study the Thermodynamics of the Witten black hole \cite{Perry, McGuigan,Nappi}. For our black hole solutions presented below,  both the on-shell action and $\delta I$ diverge. Following the techniques developed in \cite{Davis} and generalized in \cite{Grumiller}, we apply the method of Hamilton-Jacobi \cite{Martelli} to remove the aforementioned divergences. This method is based on constructing a boundary counter-term that renders a renormalized action $\Gamma$ with suitable properties 
to be used in the approximation (\ref{aproximation partition function}). Once we have the improved action $\Gamma$ at hand, 
we compute the Thermodynamics of the the black hole configurations in the canonical ensemble using the standard approach and show that our field configurations accomplish the first law, even in the extremal case.
    
    In the remainder of this paper, in section \ref{sec2}, we present two families of $AdS$ black hole solutions in two dimensions with two scalar fields that solve the field equations derived from the corresponding scalar-tensor theory. In section \ref{sec3}, we extend the spacetime of our solutions, transforming first to Eddington-Finkelstein coordinates, and then to Kruskal coordinates designed to go through the outer and inner horizons; finally we present the Penrose diagram for these spacetimes. In section \ref{sec4}, we develop a consistent Thermodynamics for the black hole configurations presented in this work. In order to do so, first we deduce the Hawking temperature by demanding regularity of the Euclidean spacetime with periodic time. We then show in detail how to construct the counter-term  for the scalar-tensor or dilaton gravity theory in two dimensions considered here, obtaining in this manner a renormalized action. This enables us to employ the approximation (\ref{aproximation partition function}) to deduce consistent thermodynamic properties by means of the first law fulfillment. We compute the total energy $M$ for the black hole solutions employing the gravitational Hamiltonian. In section \ref{sec5} we compare our system with previously reported dilatonic models. 
    In section \ref{sec6_0} we employ holography to study the effective theory living at the boundary cut-off of our black hole spacetime solution I, the solutions to the equation of motion and the Thermodynamics. Finally, we conclude in section \ref{sec6} with some remarks.

\section{Two-dimensional scalar-tensor theory and $AdS$ black holes} \label{sec2}

We shall start by considering the following action in $1+1$ dimensions 
\begin{equation} \label{action}
    S = \int d^{2} x \sqrt{-g} \; e^{\sum_{a} \gamma_a \phi_a} \left ( R + \sum_{b,c} \beta_{bc} (\partial^{\mu} \phi_b)(\partial_{\mu} \phi_c) - 2 \Lambda \right ),
\end{equation}
where $R$ is the Ricci scalar, $g$ is the determinant of the metric, $\Lambda$ is the cosmological constant, $\phi_a$ is a set of scalar fields, $\gamma_a$ are arbitrary real constant numbers, $\beta_{bc}$ stands for a square symmetric matrix whose elements are arbitrary real constant numbers, and $a$, $b$, $c$ $= 1$, $2$,...,$n$, where $n$ is the number of scalar fields. The action (\ref{action}) has been employed in  \cite{Marikasyk} to realize, holographically,
the generalized conformal structure \cite{Jevicki,Jevicki2,Jevicki3} of theories involving multiple scalar field operators, such as the SYK theory \cite{Sachdev,Kitaev}.

The equations of motion following from this action read 
\begin{equation}\label{ec einstein}
    \begin{split}
        &R_{\mu \nu} + \sum_{a,b} \beta_{ab} \;\partial_{\mu} \phi_a \;\partial_{\nu} \phi_b - \frac{1}{2} g_{\mu \nu} \left (R + \sum_{a,b} \beta_{ab} \; \partial^{\rho} \phi_a \; \partial_{\rho} \phi_b - 2\Lambda \right )- \sum_a \gamma_a \nabla_{\mu} \partial_{\nu} \phi_a -\\ 
        &\sum_{a,b} \gamma_a \gamma_b \;\partial_{\mu} \phi_a \;\partial_{\nu} \phi_b +g_{\mu \nu} \left ( \sum_{a,b} \gamma_a \gamma_b \;\partial^{\rho}\phi_a \;\partial_{\rho} \phi_b + \sum_a \gamma_a \nabla^2 \phi_a \right )= 0,
    \end{split}
\end{equation}
%
%
\begin{equation}
            \gamma_a \left[ R+\sum_{b,c} \beta_{bc} \;\partial^\sigma \phi_b \;\partial_\sigma \phi_c- 2\Lambda \right]
        =2 \sum_{c} \beta_{ac} \left( \nabla^{2} \phi_c + \nabla^{\mu} \phi_c \sum_{d}\gamma_d \;\partial_\mu \phi_d \right),
\end{equation}        
here a slight difference is noted in comparison to \cite{Marikasyk} where the overall factor $2$ is missing in the right-hand side of this equation.

Given that the product $(\partial^{\mu} \phi_b)(\partial_{\mu} \phi_c)$ is commutative, the matrix $\beta_{bc}$ is symmetric, therefore, we can diagonalize it through an $SO(n)$ transformation and obtain the matrix with diagonal elements $\beta_b$. Thus, the action (\ref{action}) in the diagonal frame is
\begin{equation} \label{action_diagonal}
    S = \int d^{2} x \sqrt{-g} \; e^{\sum_{a} \gamma_a \phi_a} \left ( R + \sum_{b} \beta_{b} (\partial^{\mu} \phi_b)(\partial_{\mu} \phi_b) - 2 \Lambda \right ).
\end{equation}

The corresponding field equations derived from this action are
\begin{equation}\label{ec einstein diagonal}
    \begin{split}
        &R_{\mu \nu} + \sum_{a} \beta_{a} \;\partial_{\mu} \phi_a \;\partial_{\nu} \phi_a - \frac{1}{2} g_{\mu \nu} \left (R + \sum_{a} \beta_{a} \; \partial^{\rho} \phi_a \; \partial_{\rho} \phi_a - 2\Lambda \right )- \sum_a \gamma_a \nabla_{\mu} \partial_{\nu} \phi_a -\\ 
        &\sum_{a,b} \gamma_a \gamma_b \;\partial_{\mu} \phi_a \;\partial_{\nu} \phi_b +g_{\mu \nu} \left ( \sum_{a,b} \gamma_a \gamma_b \;\partial^{\rho}\phi_a \;\partial_{\rho} \phi_b + \sum_a \gamma_a \nabla^2 \phi_a \right )= 0,
    \end{split}
\end{equation}
%
%
\begin{equation}
            \gamma_a \left[ R+\sum_{b} \beta_{b} \;\partial^\sigma \phi_b \;\partial_\sigma \phi_b- 2\Lambda \right]
        =2 \beta_{a} \left( \nabla^{2} \phi_a + \nabla^{\mu} \phi_a \sum_{d}\gamma_d \;\partial_\mu \phi_d \right).
\end{equation}

We further consider a static configuration of two scalar fields  under the metric ansatz (\ref{ansatz}) and arrive at the following Einstein equations

\textbf{rr:}
\begin{equation}\label{eq rr}
    \begin{aligned}
    &\beta _{1} \phi _1'(r)^2+\beta _{2} \phi _2'(r)^2+\gamma _1  \left(\frac{2}{r}+\frac{f'(r)}{f(r)}\right)\phi _1'(r)+
    \gamma _2 \left(\frac{2}{r}+\frac{f'(r)}{f(r)}\right)\phi _2'(r)\\
    &+\frac{2 l^2 \Lambda}{r^2 f(r)}=0,
    \end{aligned}
\end{equation}

\textbf{tt:}
\begin{equation}\label{eq tt}
    \begin{aligned}
       &\gamma_1 \phi_1''(r) + \gamma_2 \phi_2''(r) +\left( \gamma_{1}^2 - \frac{\beta_{11}}{2} \right) \phi_{1}'(r)^2 + \left( \gamma_{2}^2 - \frac{\beta_{22}}{2} \right) \phi_{2}'(r)^2+\\
       &2\gamma_1 \gamma_2  \phi_1 '(r) \phi_2 '(r) + \left(\frac{1}{r}+\frac{f'(r)}{2 f(r)}\right)\left(\gamma_1\phi _1'(r)+\gamma_2 \phi _2'(r)\right)+
       \frac{l^2 \Lambda}{r^2 f(r)}=0,
    \end{aligned}
\end{equation}
and the scalar field equations
\begin{equation}\label{scalar eq1}
    \begin{split}
        &2 \beta _{1} \phi _1''(r)+\beta _{1} \gamma _1 \phi _1'(r){}^2-\beta _{2} \gamma _1 \phi _2'(r){}^2+
        2 \beta _{1} \gamma _2 \phi _1'(r)\phi _2'(r) + 2\left(\frac{2}{r}+\frac{f'(r)}{f(r)}\right) \beta _{1}\phi _1'(r)\\
        &+\frac{\gamma _1 }{r^2 f(r)}\left[ r^{2} f''(r)+ 4 r f'(r)+2 f(r)+2 l^2 \Lambda\right]=0,
    \end{split}
\end{equation}
\begin{equation}\label{scalar eq2}
    \begin{split}
        &2 \beta _{2}\phi _2''(r)-\beta _{1} \gamma _2 \phi _1'(r){}^2+\beta _{2} \gamma _2 \phi _2'(r){}^2+2 \beta _{2} \gamma _1 \phi _1'(r)\phi _2'(r) + 2 \left(\frac{2}{r}+\frac{f'(r)}{f(r)}\right) \beta _{2} \phi _2'(r)\\
        &+\frac{\gamma _2 }{r^2 f(r)}\left[ r^{2} f''(r)+4 r f'(r)+2 f(r)+2 l^2 \Lambda\right]=0.
    \end{split}
\end{equation}

\subsection{$AdS_2$ black hole solutions} \label{seccion de soluciones}

In this section we present two novel analytic $AdS_2$ black hole solutions for the previous field equations (\ref{eq rr})-(\ref{scalar eq2}). To obtain the metric solutions and the scalar field configurations that support them, we first decouple the field equations for the functions $\phi_a(r)$ and $f(r)$ by solving the Euler differential equation obtained from (\ref{scalar eq1}) and (\ref{scalar eq2}) 
    \begin{equation}
        r^{2} f''(r)+4 r f'(r)+2 f(r)+2 l^2=k,
    \end{equation}
    where $k$ is an arbitrary real constant. Consequently, we find that
the blackening factor is given by
    \begin{equation} \label{pre-blackhole I}
    f(r)=f_0-\frac{c_1}{r}+\frac{c_2}{r^{2}}, \quad \quad f_0=\frac{k}{2}-l^{2}\Lambda,
\end{equation}
where $c_1$, $c_2$ are arbitrary real constants.

{\bf Solution I.}  In order to derive this solution we manipulate algebraically the field equations. We first subtract the tt equation (\ref{eq tt}) from the scalar field ones, (\ref{scalar eq1}) and (\ref{scalar eq2}), in such a way that we eliminate the terms containing  the  factor $\phi_1'(r) \phi_2'(r)$. Secondly, we use the rr relation (\ref{eq rr}) to obtain two equations without the term $\frac{f'(r)}{f(r)}\phi_1'(r)$ that we add to get a single equation for $\phi_1$ and $\phi_2$ that we separate in the following way 
\begin{align} 
    &\gamma_1\;
    \phi_1''(r)+A\;\phi_1'(r)^2+u\frac{B}{r^2 f(r)}=0, \label{final eq1}\\  
    &\gamma_2\; \phi_2''(r)+C\; \phi_2'(r)^2+2\gamma_2\left(\frac{2}{r}+\frac{f'(r)}{f(r)}\right) \phi_2'(r)+(1-u)\frac{B}{r^2 f(r)}=0,  \label{final eq2}  
    \end{align}
where $A=\frac{\beta_1\left(\left(1+\beta_1\right)\beta_2-\gamma_2^2\right)-\beta_2 \gamma_1^2}{\left(1+\beta_1\right)\beta_2}$, $B=\frac{\left(\gamma_2^2-\beta_2\gamma_1^2 \right)k+4\beta_1\beta_2\Lambda}{\left(1+\beta_1\right)\beta_2}$, $C=\frac{\beta_2\left(1+\beta_1+\gamma_1^2\right)+\beta_1 \gamma_2^2}{1+\beta_1}$ and $u$ is an arbitrary constant such that the sum of the above relations results in the above-mentioned single equation.
In order to solve (\ref{final eq1}) and (\ref{final eq2}) for the scalar fields, we impose a condition over $k$ to get $B=0$,
\begin{equation}
    k=\frac{4\beta_1\beta_2\Lambda}{\beta_2\gamma_1^2-\gamma_2^2},
\end{equation}
obtaining in this manner the following scalar fields
    \begin{align}
        &\phi_1=c_3-\frac{\gamma_1}{A}\; log\left(r+c_4\right),\\
        &\phi_2= c_5 + \frac{\gamma_2}{C}\; log\left[\frac{\left(2f_0\; r_+-c_1\right)\left(2f_0\; r-c_1\right)}{f_0\left(c_1 r-c_2\right)r^{2}}-4f_0\; \;arctanh\left(\frac{2f_0\;r-c_1}{c_1-2f_0\; r_+}\right)+c_6\right]. \label{pre-phi2}
    \end{align}
By observing the scalar field in (\ref{pre-phi2}), we realize that in order to have a real $arctanh$ function its domain requires that
    \begin{equation}
       r_-<r<r_+,
    \end{equation}
with
\begin{equation}\label{pre-horizons}
    r_{\pm}=\frac{c_1 \pm \sqrt{c_1^{2}-4c_2 f_0}}{2f_0}.
\end{equation}
Nevertheless, relations (\ref{pre-horizons}) represent the zeroes of the function (\ref{pre-blackhole I}) indicating the locations of the black hole horizons. Therefore, in order to have a physical meaningful scalar field configuration we have to consider just one non-trivial scalar field $\phi_1(r)$, requiring $\gamma_2=0$.

After substituting $\gamma_2=0$, the restriction for $k$, the resulting scalar fields and the function $f(r)$ into the field equations (\ref{eq rr})-(\ref{scalar eq2}) we find that if $c_4=\frac{c_1}{2l^{2}\Lambda}$ and $\beta_1=0$
(which implies that $k=0$ and $f_0=-l^{2}\Lambda$) all the field equations are satisfied and therefore admit the following $AdS$ black hole solution
\begin{equation} \label{blackhole I}
    ds^2=l^2\left[-\left(1-\frac{c_1}{r}+\frac{c_2}{r^{2}}\right)r^{2}\;dt^{2}+\frac{dr^2}{ \left(1-\frac{c_1}{r}+\frac{c_2}{r^{2}}\right) \;r^2} \right].
\end{equation}
In this process we have chosen $\Lambda=-\frac{1}{l^2}$ for convenience, leading to 
\begin{equation} \label{blackening factor}
    f(r)=1-\frac{c_1}{r}+\frac{c_2}{r^{2}},
\end{equation}
in order to evidently recover the $AdS$ metric at $r\rightarrow \infty$.

Thus, the scalar field configuration that supports this metric consists of just one non-trivial scalar field 
\begin{align} \label{scalars I}
    &\phi_1(r)= c_3 + log \left( r - \frac{c_1}{2}\right), &
    &\phi_2(r)= c_5, 
\end{align}
where $c_3$, $c_5$ are arbitrary real constants.
Given that the scalar field solution $\phi_1$ was obtained keeping the constant $\gamma_1$ completely arbitrary, we rescale the scalar field so that $\gamma_1=1$.

From relations (\ref{pre-horizons}) and the choice $f_0=1$, the outer $r_+$ and inner $r_-$ horizons for this solution are located at
\begin{equation} \label{r+-1}
   r_{\pm}= \pm\sqrt{\frac{c_1^2}{4}-c_2}+\frac{c_1}{2}.
\end{equation}

In order to preserve the signature of the metric, that is $f(r)>0$, and for the scalar field to be well-behaved for $r> r_+> 0$, one of the following two conditions is required
\begin{align} \label{caso1i}
    & c_1 > 0,  & \frac{c_{1}^2}{4} &\geq c_2,  & r &> \sqrt{\frac{c_1^2}{4}-c_2}+\frac{c_1}{2},\\
    &c_1 \leq 0,  &  c_2 &\leq 0,  & r &> \sqrt{\frac{c_1^2}{4}-c_2}+\frac{c_1}{2}. \label{caso1ii}
\end{align}

\textbf{Solution II.} Provided that $\Lambda=0$,  $\beta_{1}=-\beta_{2}$, the field equations (\ref{eq rr})-(\ref{scalar eq2}) admit the same black hole solution (\ref{blackhole I}), but for a different scalar field configuration
\begin{align} \label{scalars II}
    &\phi_1(r)= c_3+  log\left[ \left(r^{2}-c_1 r+c_2\right)^{\sigma_1}\right], &
    &\phi_2(r)= c_4 + log\left[ \left(r^{2}-c_1 r+c_2\right)^{\sigma_2}\right],
\end{align}
where $c_3$, $c_4$ are arbitrary real constants, $\sigma_1=\frac{1}{2\beta_{2}}$,  $\sigma_2=-\frac{1}{2\beta_{2}}$. 
 For this solution it is required that $\gamma_1=\gamma_2$ and as $\gamma_2$ remains completely arbitrary we rescale the scalar fields so that we have $\gamma_2=1$.

Again, in order to preserve  the metric signature and for the scalar fields to be well-defined for $r>r_+>0$, one of the conditions (\ref{caso1i}) or (\ref{caso1ii}) needs to be fulfilled.

{\bf Extremal case.} We would like to note that the extremal configuration  
\begin{equation} \label{extreme blackhole}
    ds^2=l^2\left[-\left(1-\frac{c_1}{2r}\right)^2 r^{2}\;dt^{2}+\frac{dr^2}{ \left(1-\frac{c_1}{2r}\right)^2 \;r^2} \right],
\end{equation}
of our black hole solutions with metric (\ref{blackhole I}), is obtained by imposing the following relation between the constants of integration
\begin{equation}\label{extrem relation}
    c_2=\frac{c_1^2}{4}.
\end{equation}
As a result of this identification the horizon, $r_H=\frac{c_1}{2}$, emerges as the union of the outer and inner horizons. In order to clarify this point, we study the spacetime structure of our solutions in section \ref{sec3}.

An important remark is that the solution (\ref{blackhole I}) reproduces, as a particular case when $c_1=0$, the two-dimensional AdS black hole configuration in the group of the $a$-$b$ family of solutions in \cite{Katanaev, Grumiller}, for $b=1$.

\textbf{Constant curvature.}  Finally, we would like  to highlight that for the solutions presented above, the curvature scalar derived from the ansatz (\ref{ansatz})
\begin{equation}
    R=g^{\mu \nu} R_{\mu \nu}= -\frac{r^2 f''(r)+ 4 r f'(r)+2 f(r)}{l^2},
\end{equation}
 results in a constant and negative quantity, an intrinsic property of the $AdS_2$ spacetime
\begin{equation} \label{curvature}
    R= -\frac{2}{l^2},
\end{equation}
 upon substitution of the metric function (\ref{blackening factor}). Moreover, we would like to note that, even though, the field configuration II has a null cosmological constant, the resulting spacetime solution possesses the above constant and negative curvature.

\begin{figure}
\centering
\includegraphics[width=0.5\linewidth, height=5.5cm]{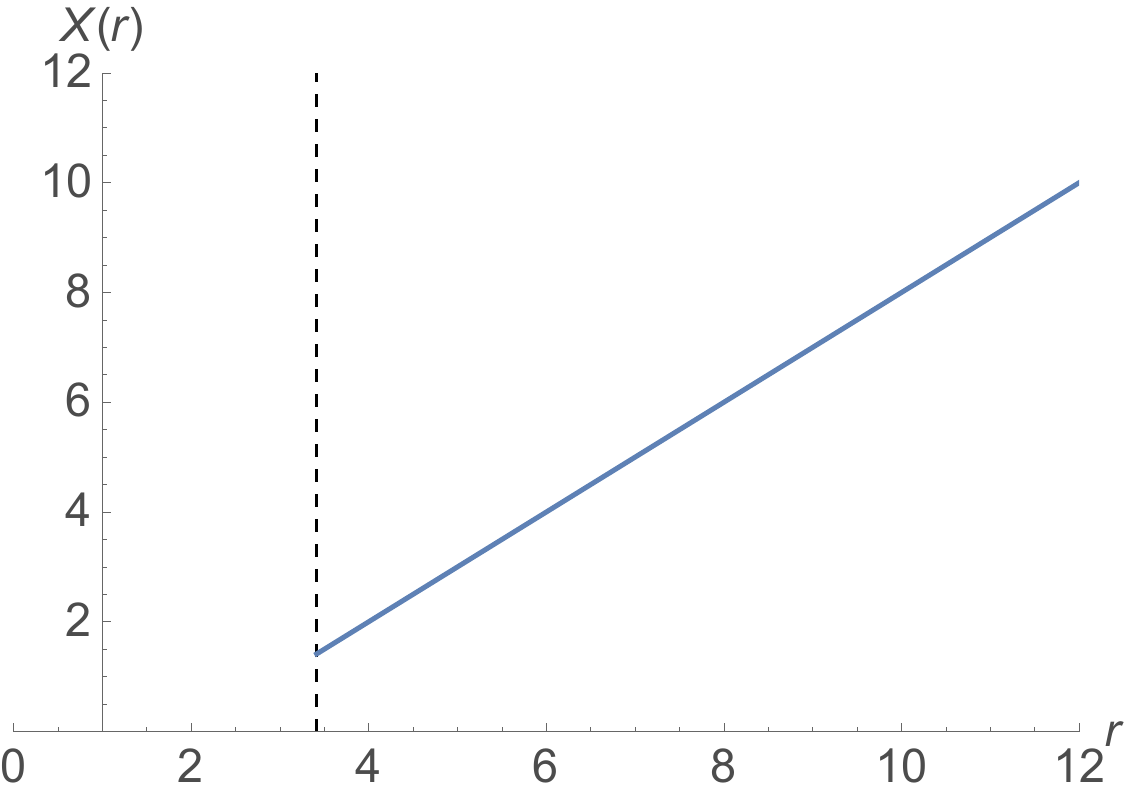}

\caption{Graphical representation of an example of the dilaton field $X(r) \equiv e^{ \phi_1+\phi_2}=r-\frac{c_1}{2}$ in the solution I. We employ the following particular values: $c_3=c_5=0$, $c_1=4$, $c_2=2$. We observe a finite value of the dilaton field $X(r)$ at the horizon $r_+=\sqrt{2}+2$, represented here with the dashed vertical line, and its asymptotically singular behavior.}
\label{img:dilaton field}
\end{figure}

{\bf Dilaton field.} At this point, it is helpful to define the dilaton field $X(r)$ as the exponential function in the action (\ref{action_diagonal})
\begin{equation}
   X(r)= e^{\sum_a \phi_a}, 
\end{equation}
here $a=1,2$, see Figure \ref{img:dilaton field}. 
As we will corroborate in the following sections, this quantity is essential in two-dimensional dilaton gravity models; for instance, it is associated with a conserved charge, defining in this way the corresponding dilatonic potential and determining the thermodynamical properties of the black holes. In addition, the value of the dilaton field at the horizon is encountered to define the entropy, giving the dilaton a relevant physical significance \cite{Perry,Nappi,Gegenberg,Davis,Grumiller,Grumiller4}.

-\subsection{$AdS_2$ geometry}

As commented above, the $AdS_2$ geometry emerges as the near-horizon limit of, for instance, the $4D$ extremal  or near-extremal RN black hole; for a pair of nice reviews see \cite{Sarosi,Trunin}. In this example it is usually  shown that, with an appropriate change of variables, the product of $AdS_2$ space and a two-dimensional sphere $S_2$, with radius equal to the dilaton, is obtained near the event horizon. That suitable change of coordinates makes use of a small parameter that sizes the separation from the horizon; eventually this parameter is set to zero as a part of the near-horizon limit.

As stated in the last section, if the relation (\ref{extrem relation}) is met in our two-dimensional solutions, the black holes become extremal. In this scenario we can implement the following change of coordinates 
\begin{equation} \label{our change of coord}
    \begin{aligned}
    &r=\frac{c_1}{2}\left(1+\Tilde{r}\right)  & &\text{and} & &t=\frac{2}{c_1}\; \Tilde{t},
    \end{aligned}
\end{equation}
where there is no near-horizon parameter and the $\Tilde{r}$ coordinate has its origin at the horizon $r_H=\frac{c_1}{2}$.  

Employing the change of coordinates (\ref{our change of coord}) in the metric describing the extreme black hole solution (\ref{extreme blackhole}), the resulting geometry description is given by the metric of the two-dimensional anti-de Sitter spacetime $AdS_2$
\begin{equation} \label{AdS2 metric}
    ds^2=l^2 \left(-\Tilde{r}^2\;d\Tilde{t}^2+\frac{d\Tilde{r}^2}{\Tilde{r}^2}\right).
\end{equation}

It is important to note that, the fact that there is no need to use a parameter that impose the near-horizon validity of the change of coordinates, tells us that we have an $AdS_2$ spacetime in all the region $r_H<r$. However, we also point out that, despite the constant curvature of our solutions, the global causal structure of the manifold, that we study in detail in the following section, reveals their black hole nature; see for instance \cite{Lemos} for another example of this kind.

\section{Black hole global causal structure} \label{sec3}

Employing the coordinates ($t$,$r$), the components of the metric (\ref{blackhole I}) have singularities at the outer and inner horizons. Therefore, in order to extend the spacetime trough this surfaces, we need to construct suitable coordinate patches.

In the case of our solutions, for radial null curves, the coordinates $t$ and $r$ are related in the following way 
\begin{equation} \label{t from r}
    t=\pm  \frac{1}{\sqrt{c_1^{2}-4 c_{2}}}\;\log \left|\frac{\sqrt{c_{1}^{2}-4 c_{2}}+c_{1}-2 r}{\sqrt{c_{1}^{2}-4 c_{2}}-c_{1}+2 r}\right|+\text{constant},
\end{equation}
where the upper/lower sign refers to null curves in the direction of increasing/decreasing $r$ (outgoing/ingoing light rays). 
From relation (\ref{t from r}) we define the coordinate 
\begin{equation} \label{r star}
    r^*= \frac{1}{\sqrt{c_1^{2}-4 c_{2}}}\;\log \left|\frac{\sqrt{c_{1}^{2}-4 c_{2}}+c_{1}-2 r}{\sqrt{c_{1}^{2}-4 c_{2}}-c_{1}+2 r}\right|,    
\end{equation}
such that $ t=\pm r^* + \text{constant}$.

Now we introduce the null coordinates  
\begin{equation} \label{u and v}
u=t-r^* \quad   \text{and}\quad v=t+r^*,
\end{equation}
  which are properly adapted to the description of null geodesics. It is easy to verify that ingoing null geodesics are described by $v=\text{constant}$ while the outgoing ones obey $u=\text{constant}$. We can use the original coordinate $r$ and replace $t$ with the coordinate $v$ or $u$. For example, if we choose the $(v,r)$ coordinate system, known as ingoing Eddington-Finkelstein coordinates, the metric (\ref{blackhole I}) takes the form
\begin{equation}
        ds^{2}=-r^{2}\left(1-\frac{c_{1}}{r}+\frac{c_{2}}{r^{2}}\right)dv^2+2dvdr.
\end{equation}
We can verify that radial null curves satisfy the following conditions:
\begin{equation}
        \frac{dv}{dr}=\left\{ \begin{array}{c l} 0 & \text{ingoing} \\\frac{2}{r^{2}\left(1-\frac{c_{1}}{r}+\frac{c_{2}}{r^{2}}\right)} & \text{outgoing,} \end{array}\right.
\end{equation}
from which we see that the light cones remain well-behaved at $r_+$ and $r_-$. Furthermore, given that the function $f(r)=1-\frac{c_{1}}{r}+\frac{c_{2}}{r^{2}}<0$ for $r_-<r<r_+$ all future directed paths of null or timelike particles are in the direction of decreasing r. This is not the case for the region $0<r<r_-$ where $f(r)>0$, consequently, the future directed paths are not necessarily pointing inwards. Below we study these and other properties of the causal structure of our solutions by means of Kruskal coordinates.

\subsection{Kruskal extension}
From definitions (\ref{r star}) and (\ref{u and v}), we see that the surface $r=r_+$ is found to be at $u=\infty$ or $v=-\infty$, so it is a good choice to construct the following null Kruskal coordinates
\begin{align} \label{coord Kruskal}
&U_{+}=\mp e^{-\kappa_{+} u}, \quad \quad V_{+}=e^{\kappa_{+} v},
\end{align}
where the upper sign in $U_+$ is used for $r>r_+$ and the lower sign for $r_-<r<r_+$; the quantity
\begin{equation} \label{surface grav}
\kappa_{+} \equiv \frac{r_{+}^{2}}{2} f^{\prime}\left(r_{+}\right)= \frac{1}{2} \sqrt{c_1^2-4c_2}, 
\end{equation}
is known as surface gravity\footnote{Given that we are dealing with a static spacetime, the event horizon $r_+$ is a Killing horizon where the Killing vector representing time translations  $\xi^\alpha=\left(\partial_t\right)^\alpha$ becomes null. We can associate to this Killing horizon the quantity $\kappa$ given by the relation
\begin{equation}
    \kappa^{2}=-\frac{1}{2}\left(\nabla^{\beta} \xi^{\alpha}\right)\left(\nabla_{\beta} \xi_{\alpha}\right).
\end{equation}
When evaluated at $r_+$, $\kappa$ is known as the surface gravity $\kappa_+ 
$.}. Here the prime denotes derivatives with respect to $r$ and we have made use of relation (\ref{r+-1}).

In order to have a metric with one timelike coordinate and one spacelike we define the following Kruskal coordinates
 \begin{align} \label{T y R}
       &T_+=\frac{1}{2}\left(V_+ + U_+\right), &R_+=\frac{1}{2}\left(V_+ - U_+\right),
   \end{align}
in terms of which the metric (\ref{blackhole I}) becomes
\begin{equation} \label{metrica T y R}
    d s^2 = 
    \frac{4}{\left(T_{+}^2-R_{+}^2+1\right)^2} \left(-dT_+^2+dR_+^2\right),
\end{equation}
where we used the following identities 
\begin{align}\label{T R de r}
    T_+^2-R_+^2&=U_+V_+=
    1- \frac{2\left(2r-c_1\right)}{2r-c_1+\sqrt{c_{1}^2-4c_2}}.
\end{align}

The form of the metric (\ref{metrica T y R}) tells us that our solutions are conformally equivalent to flat spacetime.

Using equation (\ref{T R de r}) to evaluate the metric (\ref{metrica T y R}) at  $r_+$ 
\begin{equation}
    ds^2= 4 \left(-dT_+^2+dR_+^2\right),
\end{equation}
we appreciate the nonsingular nature of the outer horizon.

Some important remarks for this coordinates are
\begin{itemize}
    \item From (\ref{metrica T y R}) we see that radial null curves look like they do in two-dimensional flat spacetime, $T_+=\pm R_+ + \text{constant}$. In particular, from (\ref{T R de r}) we see that the outer horizon $r_+$ is described by the straight lines $T_+=\pm R_+$.
    \item   From equation (\ref{T R de r}) we realize that $r=\text{constant}$ curves are described now by hyperbolae $ T_+^2-R_+^2=\text{constant}$. In particular we see that the $AdS_2$ boundary $r \rightarrow \infty$ is depicted by  $T_+^{2}-R_+^{2}\rightarrow -1$.
    \item The range for these coordinates is $ -\infty<T_+<\infty$, \ \ $R_+^2< T_+^2+1$.
    \item By virtue of equations (\ref{r+-1}) and (\ref{T R de r}) we see that the inner horizon $r_-$ is located at $T_+^2-R_+^2 \rightarrow \infty$, consequently we need a new set of Kruskal coordinates to extend the spacetime in that direction.
\end{itemize}

\begin{figure}
    \centering
    \includegraphics[scale=.4]{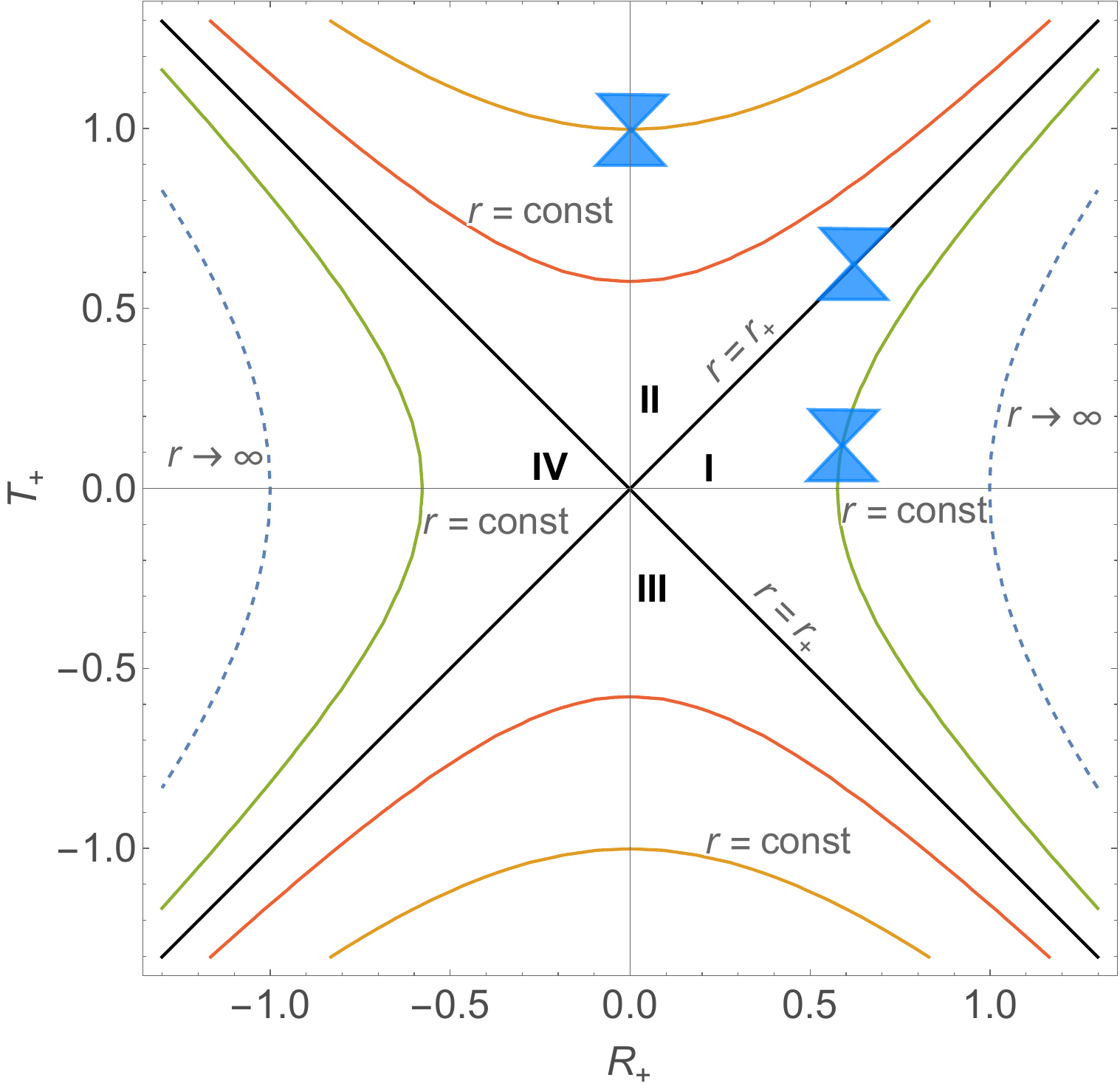}
    \caption{Kruskal diagram in coordinates ($T_+$, $R_+$).}
    \label{kruskal diagram}
\end{figure}

We can draw the Kruskal diagram in the plane $T_+-R_+$, as shown in Figure \ref{kruskal diagram}, that illustrates the regions outside (I and IV) and inside (II and III) $r_+$ up to an arbitrary cutoff surface $r_1>r_-$. We see that for $r_-<r<r_+$ all null and timelike future directed paths are in the direction of decreasing $r$ revealing that the null surface $r_+$ is an event horizon.

As for the outer horizon $r_+$ case, we construct a new set of Kruskal coordinates $T_-$ and $R_-$ adapted to the inner horizon $r_-$. In this case we define the null coordinates 
\begin{align} \label{coord Kruskal -}
&U_{-}=\mp e^{-\kappa_{-} u}, \quad V_{-}=-e^{\kappa_{-} v},
\end{align}
where the upper sign in $U_-$ is used for $r>r_-$ and the lower sign refers to $r<r_-$ and $\kappa_{-}=\frac{r_{-}^{2}}{2} f^{\prime}\left(r_{-}\right)$. The surface $r_-$ is located at $v\rightarrow \infty$ or $u\rightarrow -\infty$. As in the previous case we define the following Kruskal coordinates
 \begin{align} \label{T y R -}
       &T_{-}=\frac{1}{2}\left(V_- + U_-\right), &R_-=\frac{1}{2}\left(V_- - U_-\right),
   \end{align}
in terms of which the metric (\ref{blackhole I}) becomes
\begin{equation} \label{metrica T y R -}
    d s^2 =
    \frac{4}{\left(T_{-}^2-R_{-}^2+1\right)^2} \left(-dT_{-}^2+dR_{-}^2\right),
\end{equation}
where we used the identities
\begin{align}\label{T R de r -}
    T_{-}^2-R_{-}^2&= U_-V_-=
    1-\frac{2 \left(2r-c_1\right)}{2r-c_1-\sqrt{c_{1}^2-4c_2}}.
\end{align}

The former equation (\ref{T R de r -}) becomes identically zero when evaluated at the inner horizon $r_-$ (\ref{r+-1}). Using this fact in equation (\ref{metrica T y R -}), we realize that the metric is manifestly regular at $r_-$, i.e. $ds^2= 4 \left(-dT_-^2+ dR_-^2\right)$.  

\begin{figure}
    \centering
    \includegraphics[scale=.4]{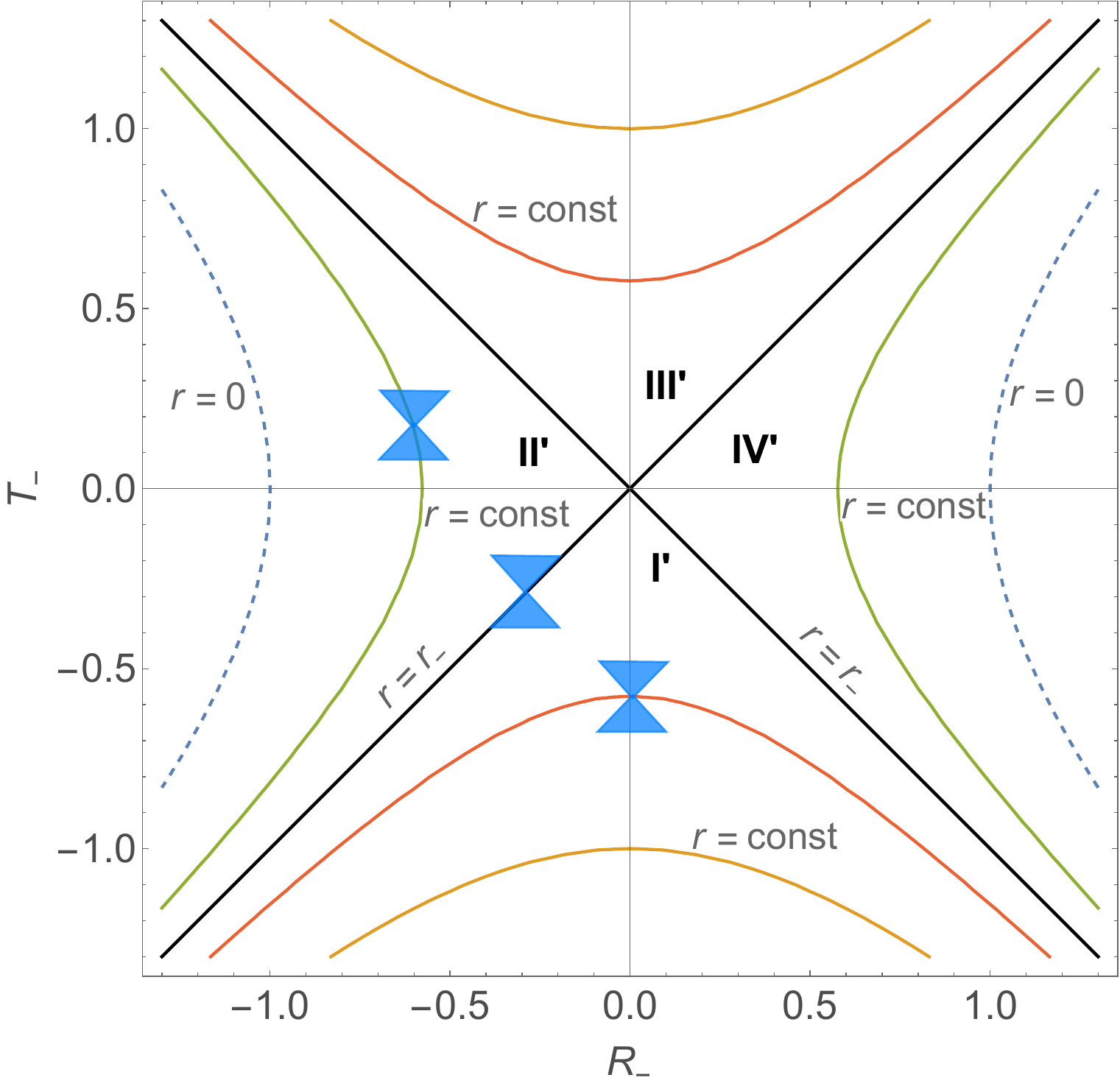}
    \caption{Kruskal diagram for coordinates ($T_-$, $R_-$).}
    \label{kruskal diagram 2}
\end{figure}

With this knowledge at hand, we draw the Kruskal diagram, as illustrated in Figure \ref{kruskal diagram 2}, for the region $0<r<r_1$ that contains the inner horizon $r=r_-$. 
A remarkable feature is that, for $r<r_-$, any surface $r=\text{constant}$ is a timelike surface, including the singularity $r=0$. This is due to the re-interpretation of $r$ as a spacelike coordinate because $f>0$ for $r<r_-$. Because of this, the singularity can be avoided, if so decided, by observers moving within the region $0<r<r_-$ of the black hole, since there, the light cone's direction allows for this kind of motion. We illustrate this interesting feature in the following section.

\subsection{Penrose diagram}

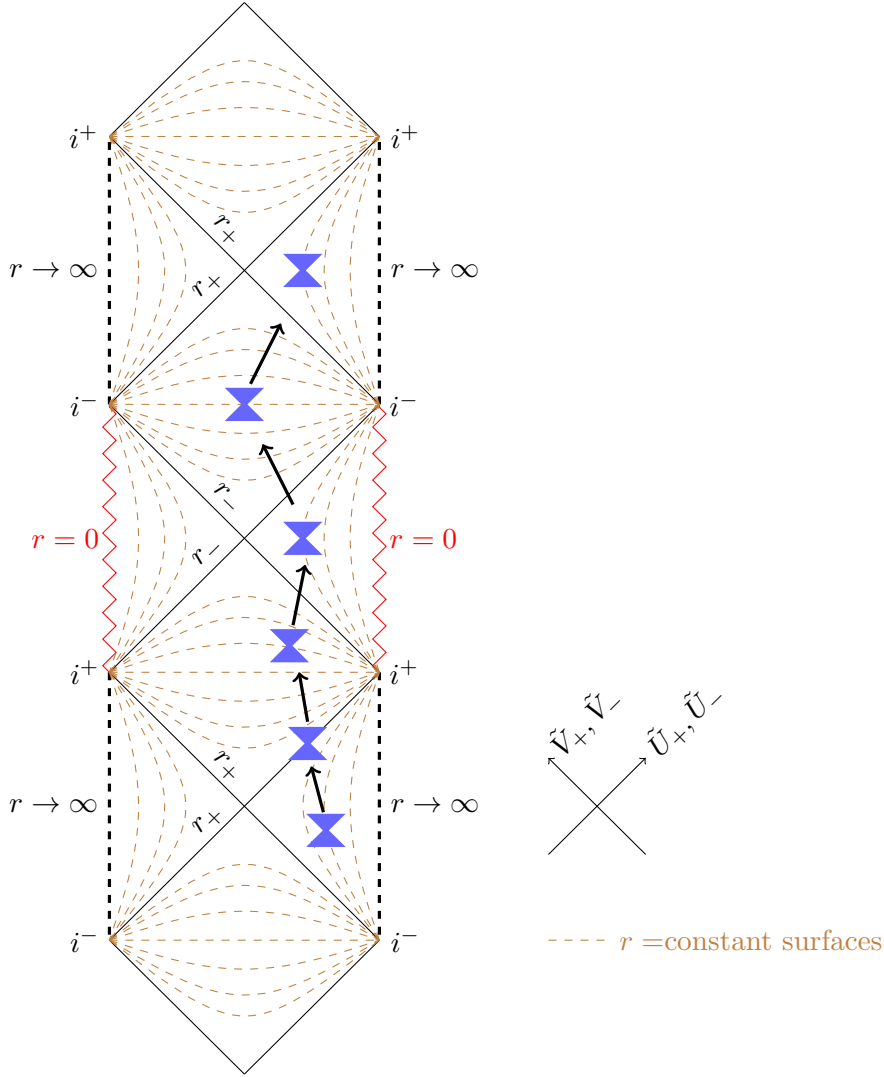
\begin{figure}
\centering
\begin{tikzpicture}[scale=1.6] 

\draw (-1.11,-1.11) -- (0,-2.22);
\draw (1.11,-1.11) -- (0,-2.22);

\draw (-1.11,1.11) node[anchor=east]{$i^+$} --node[pos=0.38,
above=-2.5,
rotate=-45]{$r_+$} 
(1.11,-1.11);
\draw (-1.11,-1.11) node[anchor=east]{$i^-$} --node[pos=0.42,
above=-2.5,
rotate=45]{$r_+$} 
(1.11,1.11) node[anchor=west]{$i^+$};
\draw[dashed,very thick] (-1.11,1.11) --node[anchor=east]{$r \rightarrow \infty$} (-1.11,-1.11);
\draw[dashed,very thick] (1.11,1.11) --node[anchor=west]{$r \rightarrow \infty$} (1.11,-1.11) node[anchor=west]{$i^-$};

\draw (-1.11,1.11) --node[pos=0.42,
above=-2.5,
rotate=45]{$r_-$} (1.11,3.33);
\draw (-1.11,3.33) --
node[pos=0.38,above=-2.5,rotate=-45]{$r_-$} (1.11,1.11);

\draw [red, decorate, decoration=zigzag](-1.11,1.11) --node[anchor=east]{$r=0$} (-1.11,3.33) ; 
\draw [red, decorate, decoration=zigzag](1.11,1.11) --node[anchor=west]{$r=0$} (1.11,3.33);

\draw (-1.11,3.33) node[anchor=east]{$i^-$} --node[pos=0.42,
above=-2.5,
rotate=45]{$r_+$}(1.11,5.55);
\draw (-1.11,5.55)--node[pos=0.38,
above=-2.5,
rotate=-45]{$r_+$}(1.11,3.33)node[anchor=west]{$i^-$};
\draw[dashed,very thick] (-1.11,3.33) --node[anchor=east]{$r \rightarrow \infty$} (-1.11,5.55)node[anchor=east]{$i^+$};
\draw[dashed,very thick] (1.11,3.33) --node[anchor=west]{$r \rightarrow \infty$} (1.11,5.55)node[anchor=west]{$i^+$};

\draw (-1.11,5.55) -- (0,6.66);
\draw (1.11,5.55) -- (0,6.66);


\draw[dashed,color=brown,domain=0.001:1.57,rotate=45] plot(\x,{pi/180*atan(-.5/tan(\x r))});
\draw[dashed,color=brown, domain=0.001:1.57,rotate=45] plot(\x,{pi/180*atan(-.25/tan(\x r))});
\draw[dashed,color=brown,domain=0.001:1.57,rotate=45] plot(\x,{pi/180*atan(-.125/tan(\x r))});
\draw[dashed,color=brown](-1.11,-1.11) -- (1.11,-1.11);

\draw[dashed,color=brown,domain=0.001:1.57,rotate=-135] plot(\x,{pi/180*atan(-.5/tan(\x r))});
\draw[dashed,color=brown,domain=0.001:1.57,rotate=-135] plot(\x,{pi/180*atan(-.25/tan(\x r))});
\draw[dashed,color=brown,domain=0.001:1.57,rotate=-135] plot(\x,{pi/180*atan(-.125/tan(\x r))});

\draw[dashed,color=brown,domain=0.001:1.57,rotate=45] plot(\x,{pi/180*atan(.5/tan(\x r))});
\draw[dashed,color=brown,domain=0.001:1.57,rotate=45] plot(\x,{pi/180*atan(.25/tan(\x r))});
\draw[dashed,color=brown,domain=0.001:1.57,rotate=45] plot(\x,{pi/180*atan(.125/tan(\x r))});
\draw[dashed,color=brown](-1.11,1.11) -- (1.11,1.11);

\draw[dashed,color=brown,domain=0.001:1.57,rotate=-45] plot(\x,{pi/180*atan(-.125/tan(\x r))});
\draw[dashed,color=brown,domain=0.001:1.57,rotate=-45] plot(\x,{pi/180*atan(-.25/tan(\x r))});
\draw[dashed,color=brown,domain=0.001:1.57,rotate=-45] plot(\x,{pi/180*atan(-.5/tan(\x r))});

\begin{scope}[shift={(0,-2.22)}]
\draw[dashed,color=brown,domain=0.001:1.57,rotate=45] plot(\x,{pi/180*atan(.5/tan(\x r))});
\draw[dashed,color=brown,domain=0.001:1.57,rotate=45] plot(\x,{pi/180*atan(.25/tan(\x r))});
\draw[dashed,color=brown,domain=0.001:1.57,rotate=45] plot(\x,{pi/180*atan(.125/tan(\x r))});
\end{scope}


\begin{scope}[shift={(0,2.22)}]
\draw (0,1.11)[dashed,color=brown,domain=0.001:1.57,rotate=45] plot(\x,{pi/180*atan(-.5/tan(\x r))});
\draw[dashed,color=brown, domain=0.001:1.57,rotate=45] plot(\x,{pi/180*atan(-.25/tan(\x r))});
\draw[dashed,color=brown,domain=0.001:1.57,rotate=45] plot(\x,{pi/180*atan(-.125/tan(\x r))});

\draw[dashed,color=brown,domain=0.001:1.57,rotate=-135] plot(\x,{pi/180*atan(-.5/tan(\x r))});
\draw[dashed,color=brown,domain=0.001:1.57,rotate=-135] plot(\x,{pi/180*atan(-.25/tan(\x r))});
\draw[dashed,color=brown,domain=0.001:1.57,rotate=-135] plot(\x,{pi/180*atan(-.125/tan(\x r))});

\draw[dashed,color=brown,domain=0.001:1.57,rotate=45] plot(\x,{pi/180*atan(.5/tan(\x r))});
\draw[dashed,color=brown,domain=0.001:1.57,rotate=45] plot(\x,{pi/180*atan(.25/tan(\x r))});
\draw[dashed,color=brown,domain=0.001:1.57,rotate=45] plot(\x,{pi/180*atan(.125/tan(\x r))});
\draw[dashed,color=brown](-1.11,1.11) -- (1.11,1.11);

\draw[dashed,color=brown,domain=0.001:1.57,rotate=-45] plot(\x,{pi/180*atan(-.125/tan(\x r))});
\draw[dashed,color=brown,domain=0.001:1.57,rotate=-45] plot(\x,{pi/180*atan(-.25/tan(\x r))});
\draw[dashed,color=brown,domain=0.001:1.57,rotate=-45] plot(\x,{pi/180*atan(-.5/tan(\x r))});
\end{scope}

\begin{scope}[shift={(0,4.44)}]
\draw [dashed,color=brown,domain=0.001:1.57,rotate=45] plot(\x,{pi/180*atan(-.5/tan(\x r))});
\draw[dashed,color=brown,domain=0.001:1.57,rotate=45] plot(\x,{pi/180*atan(-.25/tan(\x r))});
\draw[dashed,color=brown,domain=0.001:1.57,rotate=45] plot(\x,{pi/180*atan(-.125/tan(\x r))});

\draw[dashed,color=brown,domain=0.001:1.57,rotate=-135] plot(\x,{pi/180*atan(-.5/tan(\x r))});
\draw[dashed,color=brown,domain=0.001:1.57,rotate=-135] plot(\x,{pi/180*atan(-.25/tan(\x r))});
\draw[dashed,color=brown,domain=0.001:1.57,rotate=-135] plot(\x,{pi/180*atan(-.125/tan(\x r))});

\draw[dashed,color=brown,domain=0.001:1.57,rotate=45] plot(\x,{pi/180*atan(.5/tan(\x r))});
\draw[dashed,color=brown,domain=0.001:1.57,rotate=45] plot(\x,{pi/180*atan(.25/tan(\x r))});
\draw[dashed,color=brown,domain=0.001:1.57,rotate=45] plot(\x,{pi/180*atan(.125/tan(\x r))});
\draw[dashed,color=brown](-1.11,1.11) -- (1.11,1.11);

\draw[dashed,color=brown,domain=0.001:1.57,rotate=-45] plot(\x,{pi/180*atan(-.125/tan(\x r))});
\draw[dashed,color=brown,domain=0.001:1.57,rotate=-45] plot(\x,{pi/180*atan(-.25/tan(\x r))});
\draw[dashed,color=brown,domain=0.001:1.57,rotate=-45] plot(\x,{pi/180*atan(-.5/tan(\x r))});
\end{scope}

\begin{scope}[shift={(0,6.66)}]
\draw[dashed,color=brown,domain=0.001:1.57,rotate=-45] plot(\x,{pi/180*atan(-.125/tan(\x r))});
\draw[dashed,color=brown,domain=0.001:1.57,rotate=-45] plot(\x,{pi/180*atan(-.25/tan(\x r))});
\draw[dashed,color=brown,domain=0.001:1.57,rotate=-45] plot(\x,{pi/180*atan(-.5/tan(\x r))});
\end{scope}


\draw[blue!60, ultra thick,fill=blue!60] (.55,-.32) -- (.79,-.32) -- (.67,-.20) -- cycle;
\draw[blue!60, ultra thick,fill=blue!60] (.55,-.08) -- (.79,-.08) -- (.67,-.20) -- cycle;

\draw[very thick, ->](.65,-.05) -- (.55,.33);

\draw[blue!60,ultra thick,fill=blue!60] (.40,.40) -- (.64,.40) -- (.52,.52) -- cycle;
\draw[blue!60, ultra thick,fill=blue!60] (.40,.64) -- (.64,.64) -- (.52,.52) -- cycle;

\draw[very thick, ->](.52,.7) -- (.45,1.11);


\draw[blue!60, ultra thick,fill=blue!60] (.49,1.45) -- (.25,1.45) -- (.37,1.33) -- cycle;
\draw[blue!60, ultra thick,fill=blue!60] (.37,1.33) -- (.25,1.21) -- (.49,1.21) -- cycle;

\draw[very thick, ->](.40,1.5) -- (.50,2);

\draw[blue!60, ultra thick,fill=blue!60] (.36,2.34) -- (.60,2.34) -- (.48,2.22) -- cycle;
\draw[blue!60, ultra thick,fill=blue!60] (.36,2.10) -- (.60,2.10) -- (.48,2.22) -- cycle;

\draw[very thick, ->](.4,2.5) -- (.15,3);

\draw[blue!60, ultra thick,fill=blue!60] (-.12,3.45) -- (.12,3.45) -- (0,3.33) -- cycle;
\draw[blue!60, ultra thick,fill=blue!60] (0,3.33) -- (-.12,3.21) -- (.12,3.21) -- cycle;

\draw[very thick, ->](.05,3.5) -- (.3,4);

\draw[blue!60, ultra thick,fill=blue!60] (.36,4.56) -- (.60,4.56) -- (.48,4.44) -- cycle;
\draw[blue!60, ultra thick,fill=blue!60] (.36,4.32) -- (.60,4.32) -- (.48,4.44) -- cycle;

\begin{scope}[shift={(0,-1.11)}]
\draw[<-] (2.5,1.51)node[anchor=west,rotate=45]{$\tilde{V}_+,\tilde{V}_-$} -- (3.3,.71);
\draw[->] (2.5,.71)-- (3.3,1.51)node[anchor=west,rotate=45]{$\tilde{U}_+,\tilde{U}_-$};

\draw[color=brown,dashed](2.5,0) -- (3,0)node[anchor=west]{$r=$constant surfaces};
\end{scope}

\end{tikzpicture}
\caption{Conformal diagram for our $AdS_2$ black holes. The spacetime boundary $r \rightarrow \infty$ is represented by the dashed vertical straight lines and the singularity $r=0$ is depicted by the zigzag vertical ones. Here we illustrate a particular motion through a timelike path.}
\label{Penrose diagram}
\end{figure}

In order to illustrate our two-dimensional spacetimes in compact form, we construct their Penrose diagram, shown in Figure \ref{Penrose diagram}.

We employ the coordinate transformation 
\begin{equation} \label{conformal trans}
    \tilde{U}_{\pm}=arctan(U_{\pm})\;\; \text{and} \; 
    \; \tilde{V}_{\pm} = arctan(V_{\pm}),
\end{equation}
over the null Kruskal coordinates (\ref{coord Kruskal}) and (\ref{coord Kruskal -}). The relations (\ref{T y R}), (\ref{T y R -}) and the range of the coordinates $\left(T_+,R_+\right)$, which is the same for $\left(T_-,R_-\right)$, are employed to deduce that the range of the coordinates $\left(U_+,V_+\right)$ and $\left(U_-,V_-\right)$ is given by
\begin{equation}
    -\infty < U_{\pm} < \infty \;\; \text{and} \;\; -1 < U_{\pm} V_{\pm}.
\end{equation}
Taking into account this last relation and the transformation (\ref{conformal trans}) we easily realize that the $\left(\tilde{U}_+,\tilde{V}_+\right)$ and $\left(\tilde{U}_-,\tilde{V}_-\right)$ coordinate domains are defined by the intersection of
\begin{equation}
    -\frac{\pi}{2} < \tilde{U}_{\pm} < \frac{\pi}{2}, \;\;\; -\frac{\pi}{2} < \tilde{V}_{\pm} < \frac{\pi}{2}  \;\;\text{and} 
    \;\; -1 < tan\left(\tilde{V}_{\pm} \right) tan\left(\tilde{U}_{\pm}\right).
\end{equation}

From the relations (\ref{T y R}) and (\ref{T y R -}), the metric expressions (\ref{metrica T y R}) and (\ref{metrica T y R -}), and the transformation (\ref{conformal trans}), we find the metric form
\begin{equation}
    ds^2= -4 \frac{d \tilde{U}_{\pm} d\tilde{V}_{\pm}}{cos^2\left(\tilde{U}_{\pm}-\tilde{V}_{\pm}\right)}.
\end{equation}
From this equation we observe that the light cones are depicted with lines $\tilde{U}_{\pm}=$constant or $\tilde{V}_{\pm}=$constant. The outer and inner horizons $r_+$ and $r_-$ are described with the same straight lines $\tilde{U}_{\pm} \tilde{V}_{\pm}=0$ as in Kruskal coordinates. The timelike boundary $r\rightarrow \infty$ is now depicted in a finite form by straight lines between the future timelike infinity $i^+$ and the past timelike infinity $i^-$ which, in the same way, are the past and future end points of the surfaces $r=$constant. Similarly, the singularity $r=0$ is illustrated by straight lines in this conformal diagram.

We use the Penrose diagram to illustrate the motion of an observer inside these $AdS_2$ black holes, as shown in Figure \ref{Penrose diagram}. Once the observer has crossed the outer horizon $r=r_+$, when moving towards the interior of the event horizon, the only possible direction implies  decreasing $r$, in this region the  coordinate $r$ is timelike. After crossing the inner horizon $r=r_-$ the coordinate $r$ becomes spacelike and any direction is possible. If the observer decides to return to the inner horizon he will cross another copy of $r_-$. After that, the only option for the observer is to go towards the outer horizon because now the coordinate $r$ becomes once again timelike. The observer goes out, from another copy of the outer horizon, to a new asymptotically $AdS_2$ spacetime.

\section{Thermodynamics} \label{sec4}

For higher dimensional black holes, the standard Bekenstein-Hawking relation states that the entropy is always found to be one quarter of the horizon area, in Planck units
\begin{equation}
    S=\frac{A_H}{4G}.
\end{equation}
In \cite{Grumiller} it was shown that this formula also holds for the two-dimensional dilaton-gravity case with an effective Newton constant defined in terms of the dilaton field at the horizon $G_{\text{eff}}=\frac{G_2}{X_H}$, namely, 
\begin{equation} \label{entropy2d}
    S=\frac{A_H}{4G_{\text{eff}}}.
\end{equation}
To see how this relation takes place, we first recall that a sphere of radius $r$ in $d$ spatial dimensions has an area $A_d=2 \pi^{d / 2} r^{d-1} / \Gamma(d / 2)$. Then we make use of this formula in the $d\rightarrow 1$ limit to compute $A_1=2$. When considering that in one dimension the sphere consists of two disjoint points, only one of them is associated with the horizon, such that $A_H=A_1/2=1$. By substituting this result into (\ref{entropy2d}) we are led to
\begin{equation}
    S=\frac{1}{4G_{\text{eff}}}=\frac{X_H}{4G_{2}};
\end{equation}
by further setting $8\pi G_2=1$ we obtain the known result for the entropy of two-dimensional black holes \cite{Perry,Grumiller,Davis,Nappi,Gegenberg,Grumiller4}
\begin{equation}\label{EntropyG}
    S=2 \pi X_H ,
\end{equation}
a quantity that is completely determined by the event horizon. It is worth noticing that this is a universal result valid for any dilaton gravity model.

In this section, we verify this result for our solutions by employing the Euclidean treatment of quantum gravity \cite{Gibbons}. In this approach, the partition function $\mathcal{Z}$ is obtained by computing the path integral over the space of all periodic field configurations in Euclidean time. As stated above, the path integral is given by the approximation (\ref{aproximation partition function}) under certain conditions. With this in mind, we construct a renormalized action $\Gamma_{\text{reg}}$ with a regulating boundary $r=r_{\text{reg}}$ and obtain the partition function for the canonical ensemble in this way. Finally, we compute the thermodynamic properties for our two black hole configurations. We verify the results by accomplishing the quasi-local form of the first law of Thermodynamics.   

\subsection{Temperature} \label{sec 4.1}

In order to deduce the Hawking temperature,  following the approach first presented in \cite{Gibbons}, we consider regularity at the Euclidean horizon. As usual, see for example \cite{Hartnoll4}, we first Taylor expand the metric (\ref{ansatz}) near the outer horizon $r_+$ and obtain
 \begin{equation}
        ds^2=l^2\left(-r_{+}^{2}\left(r-r_+\right)f'(r_+) dt^2+\frac{dr^2}{r_+^2\left(r-r_+\right)f'(r_+)}+...\right).
    \end{equation}
Subsequently, we perform the Wick rotation $t \rightarrow -i t=t_E$, and carrying out the change of coordinates
\begin{align}\label{rho y eta}
        &r=r_++\frac{r_+^2 f'(r_+)}{4l^2}\rho^2, & &t_E=\frac{2}{f'(r_+)r_+^{2}}\eta,
    \end{align}
we find the near-horizon Euclidean metric to be
 \begin{equation} \label{rindler space}
        ds^2=\rho^2 d\eta^2+d\rho^2+...,
    \end{equation}
which we identify as Euclidean space in two dimensions in polar coordinates. In order to avoid a conical singularity at the Euclidean horizon $\rho=0$ it is necessary to take into account the periodicity 
 \begin{align} \label{periodicity}
        &\eta \sim \eta+2\pi, & &\text{which means} & &t_E \sim t_E+\frac{4\pi}{f'(r_+)r_+^{2}}.
 \end{align}
 Recalling that if we have a quantum field theory with a Wick rotated periodic time, with period $\beta$, then we have a theory with finite temperature $T=\frac{1}{\beta}$, assuming $\hbar=1$. Therefore we have found that the Hawking temperature of the black hole solution considered here is
 \begin{equation}\label{hawking temperature}
        T=\frac{f'(r_+)r_+^{2}}{4\pi},
    \end{equation}
which, as we can appreciate from  (\ref{surface grav}),  is related to the surface gravity in the following way
\begin{equation}\label{hawking temperature with surface g}
        T=\frac{\kappa_+}{2\pi}=\frac{1}{4 \pi}\sqrt{c_1^2-4c_2}.
    \end{equation}
From this equality it is easy to see that the extremal configuration (\ref{extreme blackhole}) has zero temperature because of the relation (\ref{extrem relation}).

We see that the temperature (\ref{hawking temperature with surface g}) reproduces as a particular case, when $c_1=0$, the Hawking temperature for the AdS black hole of the $a$-$b$ family presented in \cite{Katanaev, Grumiller} for $b=1$.

In section \ref{sec 4.2.1}, it will be helpful to relate the Hawking temperature $T$ to a local proper temperature $T_w$ measured at an arbitrary surface $r=r_w>r_+$. Given that the Hawking temperature is established by requiring the periodicity (\ref{periodicity}) in the Euclidean time $t_E$ (or in the coordinate $t$), we can employ the Euclidean relation between $t_E$ (or $t$) and the proper time $\tau_w$ for a static observer placed at $r_w$ 
\begin{equation}
    d \tau_w^{2} = g_{tt}(r_w)\; dt_E^{2} =r_{w}^{2} f(r_w) dt^2,
\end{equation}
to obtain the redshift or Tolman relation\footnote{For asymptotically flat black holes, $g_{tt}\rightarrow 1$ when $r \rightarrow \infty$, therefore, the Hawking temperature $T$ corresponds to the local temperature $T_w$ measured by an observer at infinity.} \cite{Tolman}
\begin{equation}\label{tolman factor}
    T_w=\frac{1}{r_{w} \sqrt{f(r_{w})}} T.
\end{equation}

\subsection{$AdS_2$ black hole partition function in two dimensions} \label{sec 4.2}

From this section we consider the Euclidean version of the action (\ref{action_diagonal}) in $1+1$ dimensions
\begin{equation}\label{action with gibbons term}
    I= \int _{\mathcal{M}} d^{2} x \sqrt{g} \; X \left ( R + \sum_{b} \beta_{b} (\partial^{\mu} \phi_b)(\partial_{\mu} \phi_b) - 2 \Lambda \right ) +2\int_{\partial \mathcal{M}} d x \sqrt{\gamma}\; X\; K, 
\end{equation}
where, as stated above, $X= e^{\sum_a \phi_a}$ stands for the dilaton and $a,b=1,2$. We have added the Gibbons-Hawking-York term \cite{Gibbons,York} where $\gamma_{ij}=g_{tt}$ is the induced metric on the boundary\footnote{Because we are dealing with a one-dimensional boundary, the subscripts $i,j$ just keep track of the quantities related to the induced metric.}   $r=$constant, with $r\rightarrow \infty$, and $K$ is the trace of the extrinsic curvature or second fundamental form. 

We compute the thermodynamical properties for the two black hole solutions presented in section \ref{seccion de soluciones} employing the partition function $\mathcal{Z}$ given by the path integral weighted by the exponential of the Euclidean action $I$ \cite{Gibbons}
\begin{equation}
\mathcal{Z}=\int \mathcal{D} g \mathcal{D} \phi_a \exp \left(-\frac{1}{\hbar} I[g, \phi_a]\right),
\end{equation}
with $\mathcal{D} g$ and $\mathcal{D} \phi_a$ denoting some measure for the metric  and  the scalar fields, respectively.  

We might assume that the dominant contribution for the path integral comes from the solutions to the classical field equations, so that we can approximate 
\begin{equation} \label{approximation z}
\mathcal{Z} \sim \exp \left(-\frac{1}{\hbar} I\left[g_{c l}, \phi_{a,c l}\right]\right).
\end{equation}
 Nevertheless, in order for the assumption to be valid, it is necessary to have an action that is finite on-shell and whose variation $\delta I$ vanishes for the classical solutions $g_{cl}$ and $\phi_{a,cl}$. 
 
 To evaluate the action (\ref{action with gibbons term}), we shall incorporate an auxiliary regulator $r\leq r_{\text{reg}}$, treating the surface $r=r_{\text{reg}}$ as a finite boundary; we recover the full spacetime by taking the limit $r_{\text{reg}} \rightarrow \infty$. For instance, computing the regulated on-shell action for our solution I we arrive at
\begin{equation} \label{I_reg onshell action}
    I_{\text{reg}}=2x_0 \beta \left[\left(r_{\text{reg}}-\frac{c_1}{2}\right)^2+\frac{2\pi}{x_0} X_+ T-\left(\frac{c_1^2}{4}-c_2\right)\right],
\end{equation}
where $X_+=x_0 \sqrt{\frac{c_1^2}{4}-c_2}$ is the value of the dilaton at the horizon $r_+$, the constant $x_0=e^{c_3+c_4}$ and remember that the period $\beta=\frac{1}{T}$, with $T$ being the Hawking temperature (\ref{hawking temperature with surface g}). Note that the limit $r_{\text{reg}}\to\infty$ in equation (\ref{I_reg onshell action})
diverges for the on-shell action I. Similarly, we further verify that variations of the fields, that preserve the boundary conditions in this solution, lead to 
\begin{equation}
    \delta I=\lim_{r \rightarrow \infty} \delta I_{\text{reg}}\rightarrow \infty.
\end{equation}

Following the techniques developed in \cite{Davis,Grumiller} we apply the method of Hamilton-Jacobi \cite{Martelli} to remove the divergences. This approach enables us to construct a boundary counter-term  $I_{ct}$ that renders the action finite on-shell and that is extremized  by classical solutions of the field equations. This counter-term action is related to (\ref{action with gibbons term}) in the following way
\begin{equation} \label{relation actions}
    I= I_{ct}+ \Gamma,
\end{equation}
where the resulting renormalized action $\Gamma$ is the one we are allowed to employ in the saddle point approximation (\ref{approximation z}). The boundary integral $I_{ct}$ may depend on the fields and only on their tangential derivatives to the boundary in order for the actions $I$ and $\Gamma$ to lead to the same field equations. 

In order to obtain the counter-term, the Hamiltonian derived from the action $I$ is required to satisfy the constraint $\mathcal{H}=0$. For the action (\ref{action with gibbons term}) the associated Hamiltonian density is
\begin{equation}\label{Hamiltonian density}
\begin{split} 
 \mathcal{H}&=-4\beta_{1} \beta_{2}\left(\pi^{i j} \gamma_{i j}\right)^{2}+4\pi^{i j} \gamma_{i j} \left(\beta_{2} \pi_{\phi_{1}}+\beta_{1} \pi_{\phi_{2}}\right)
 +\left(\pi_{\phi_{1}}-\pi_{\phi_{2}}\right)^{2} +8\left(\beta_{1}+\beta_{2}\right) X^{2} \Lambda;
\end{split} 
\end{equation}
here the canonical momenta, $\pi^{ij}$ and $\pi_{\phi_a}$, conjugate to the fields are defined in terms of the change of the fields along the $r$ direction \footnote{For a thorough review of the Hamiltonian formulation for a general dilaton theory see \cite{Dyer}. }.

Varying the action with respect to the fields and evaluating it for a solution of the field equations, momenta appear as boundary terms 
\begin{equation}
    \delta I_{\text{on-shell}}=\int_{\partial \mathcal{M}} d \tau  \sqrt{\gamma}\left[\pi^{ij} \delta \gamma_{ij}+\sum_{a=1}^{2} \pi_{\phi_a} \delta \phi_a\right],
\end{equation}
in such a way that we can write them as functional derivatives of the on-shell action with respect to the fields at the boundary
\begin{equation}\label{momenta as derivatives}
\pi^{i j}=\frac{1}{\sqrt{\gamma}} \frac{\delta}{\delta \gamma_{i j}}\left(I_{\text{on-shell}}\right), \quad \pi_{\phi_a}=\frac{1}{\sqrt{\gamma}} \frac{\delta}{\delta \phi_a}\left(I_{\text{on-shell}}\right).
\end{equation}

With equation (\ref{Hamiltonian density}) and the result (\ref{momenta as derivatives}), the Hamiltonian constraint $\mathcal{H}=0$ is written as a non-linear functional differential equation for the on-shell action, the Hamilton-Jacobi equation. 

Given that the counter-term action is, in the same way, required to solve the Hamilton-Jacobi equation we must have
\begin{equation}\label{Hamilton-Jacobi Ict}
\begin{aligned}
&-4\beta_{1} \beta_{2}\left[\gamma_{i j}\; \left(\partial_{\gamma_{i j}}I_{ct}\right)\right]^{2}+4 \gamma_{i j}\; \left(\partial_{\gamma_{i j}}I_{ct}\right)\left[\beta_{2} \left(\partial_{\phi_{1}} I_{ct}\right)+\beta_{1} \left(\partial_{\phi_{2}}I_{ct}\right)\right]\\
&+\left[\left(\partial_{\phi_{1}}I_{ct}\right)-\left(\partial_ {\phi_{2}}I_{ct}\right)\right]^{2}
+8\left(\beta_{1}+\beta_{2}\right) X^2 \Lambda=0.
\end{aligned}
\end{equation}
In order to solve the above non-linear differential equation we take advantage of the symmetries that $I_{ct}$ must fulfill.
First, it must be invariant under diffeomorphisms of $\partial \mathcal{M}$, accordingly the boundary integral takes the form 
\begin{equation}
    I_{ct}=\int_{\partial \mathcal{M}} d\tau\;\sqrt{\gamma}\; \mathcal{L}_{ct}\left(\phi_1,\phi_2\right),
\end{equation}
where the scalar $\mathcal{L}_{ct}$ does not depend on tangential derivatives to the boundary because the scalar fields $\phi_a$ are invariant over time.
Secondly the action (\ref{action with gibbons term}) is invariant under the transformation 
\begin{equation} \label{B simetry}
    \begin{aligned}
    g_{tt}&\rightarrow\frac{1}{g_{tt}},\\
    \phi_1 \rightarrow \phi_1+\frac{1}{2} log\left(|g_{tt}|\right), &\quad  \phi_2 \rightarrow \phi_2+\frac{1}{2} log\left(|g_{tt}|\right).
    \end{aligned}
\end{equation}
We expect that the resulting action $\Gamma$ respects the symmetries that the action $I$ possesses, therefore $I_{ct}$ must be invariant under (\ref{B simetry}). This is achieved by taking the ansatz:
\begin{equation}\label{counter-term}
 I_{ct}=C\int_{\partial\mathcal{M}} d\tau \sqrt{g_{tt}}\; e^{\phi_1+\phi_2},
\end{equation} 
where $C$ is an arbitrary constant.

The remaining part is to substitute the above expression for $I_{ct}$ into the Hamilton-Jacobi equation (\ref{Hamilton-Jacobi Ict}), then it is straightforward to determine that
\begin{equation} \label{counterterm}
    C= 2 \;\sqrt{-2 \Lambda} \left( \frac{\beta_{1}+\beta_{2}}{2\left(\beta_{1}+\beta_{2}\right)-\beta_{1}\beta_{2}}\right)^{\frac{1}{2}}.
\end{equation}

According to (\ref{action with gibbons term}), (\ref{relation actions}) and (\ref{counter-term}) the action $\Gamma$ becomes
\begin{equation}\label{action2}
\begin{aligned}
        \Gamma &= \int _{\mathcal{M}} d^{2} x \sqrt{g} \; X \left ( R + \sum_{b} \beta_{b} (\partial^{\mu} \phi_b)(\partial_{\mu} \phi_b) - 2 \Lambda \right )
        +2\int_{\partial \mathcal{M}} d x \sqrt{\gamma}\; X\; K \\
        &- C\int_{\partial\mathcal{M}} d\tau \sqrt{\gamma}\; X.
\end{aligned}        
\end{equation}

Returning to the case of our solution I, substituting the values for $\Lambda$ and $\beta_{1}$ in (\ref{counterterm}) produces a counter-term with $C=\frac{2}{l}$. As before, employing the regulatory boundary $r=r_{\text{reg}}$, we compute the regulated on-shell action $\Gamma_{\text{reg}}$ for this case:
\begin{equation} \label{reg action}
    \Gamma_{\text{reg}}= 2 x_0 \beta\left[\left(r_{\text{reg}}-\frac{c_1}{2}\right)^2+\frac{2\pi}{x_0 } X_+ T-\left(\frac{c_1^2}{4}-c_2\right)-r_{\text{reg}} \sqrt{f(r_{\text{reg}})} \left(r_{\text{reg}}-\frac{c_1}{2}\right)\right].
\end{equation}
Removing the regulator by taking the limit
\begin{equation} \label{on-shell gamma}
    \lim_{r_{\text{reg}}\to\infty} \Gamma_{\text{reg}}=2 x_0  \beta \left[\frac{2\pi}{x_0} X_+ T-\frac{1}{2}\left(\frac{c_1^2}{4}-c_2\right)\right],
\end{equation}
we verify the finite result for the renormalized on-shell action $\Gamma$. Furthermore, we compute that all variations of the fields, preserving the boundary conditions, lead to 
\begin{equation}
    \delta \Gamma=\lim_{r_{\text{reg}} \rightarrow \infty} \delta \Gamma_{\text{reg}}=0.
\end{equation}
Considering the result (\ref{on-shell gamma}), we develop the remaining thermodynamic properties for the black hole solution I in the following section.

Equation (\ref{counterterm}) shows that there is no counter-term for the remaining solution II presented in this work, as $\Lambda=0$ in this case. However, we find that the on-shell action is an appropriate finite action, with null variation, to be used in the semi-classical approximation, allowing us to develop the thermodynamic properties presented below.

\subsection{Canonical ensemble} \label{sec 4.2.1}
In usual Thermodynamics a canonical ensemble is defined by the temperature and a variable determining the size of the system, that is to say, the volume. In \cite{York2} the author designs a system consisting of a spherical cavity, delimited by a cavity wall at radius $r$, enclosing  a black hole at the center. The canonical ensemble of such a system is defined by the local constant temperature $T_w(r)$ and the area of the cavity wall. The size of the system is not specified by spatial volume because the volume of a black hole is not defined at a constant Euclidean time. 

Following the approach consisting in enclosing a black hole in a cavity developed in \cite{Perry, Davis, Grumiller}, here we perform a similar analysis. We give a physical meaning to the surface  $r_{\text{reg}}$ by imagining that it represents the wall of a ``cavity'' $r_w$ that maintains boundary conditions. The local temperature $T_w$ measured at the wall is  given by the Tolman relationship (\ref{tolman factor}). 

In two dimensions we can construct a conserved current $j^{\mu}$ from any regular function $f(\Phi)$ of a scalar field in the following way
\begin{equation}
    j^{\mu}=\epsilon^{\mu \nu} \nabla _{\nu}f(\Phi),
\end{equation}
where $\epsilon$ is the Levi-Civita tensor in two dimensions. The associated conserved charge is
\begin{equation}
    D_w=f(\Phi_w)=\int_{\Sigma} dr \sqrt{g_{rr}} \;j_{\mu} n^{\mu},
\end{equation}
where $\Sigma$ is a surface of constant time with unitary normal vector $n^{\mu}$ and a boundary located at $r=r_w$.  Following \cite{Perry} we choose the function $f(\Phi)= X$, with $\Phi \equiv \phi_1+\phi_2$, so that we have the conserved dilaton charge
\begin{equation}\label{dilaton charge}
    D_w=X_w,
\end{equation}
where the subscript $w$ indicates us that the charge depends on the location of the wall. Thus equation (\ref{dilaton charge}) gives us the dilaton charge contained within the cavity wall $r_w$. We assign to $X_w$ an analogous role to that of the area of the cavity wall in higher dimensions.

As a result we have designed a cavity delimited by a wall $r_w$ where we keep the temperature $T_w$ and dilaton charge $X_w$ fixed, hence the approximation (\ref{approximation z}) accounts for  the partition function in the canonical ensemble
\begin{equation} \label{z with Gamma}
\mathcal{Z}(T_w, X_w)=\text{exp}\left(-\Gamma_w\right),
\end{equation}
where we have made $\hbar=1$. 

The corresponding  Helmholtz free energy $F_w$ is given by
\begin{equation}\label{free energy 1}
 F_w(T_w, X_w)= T_w\;\text{log}\mathcal{Z}=-T_w\; \Gamma_w(T_w, X_w),
\end{equation}
where again, the subscript reminds us that $F_w$ is the Helmholtz free energy for the system inside the wall $r=r_w$. 

On the other hand, the first law of Thermodynamics corresponding to this canonical ensemble reads
\begin{equation}\label{first law0}
    dE_w=T_w dS_w - \psi_w dX_w,
\end{equation}
where $E_w$ is the internal energy, $S_w$ is the entropy and $\psi_w$ is the chemical potential associated with the dilaton charge, the minus sign is intended to preserve the analogy with pressure in standard Thermodynamics. As usual, see for instance \cite{Reif}, from (\ref{first law0}) and the Legendre transformation 
\begin{equation}\label{Legendre trans}
    F_w(T_w,X_w) = E_w(S_w,X_w) - T_w S_w,
\end{equation}
we arrive to the equivalent formulation
\begin{equation}
    dF_w=-S_w dT_w-\psi_w dX_w,
\end{equation}
which in turn defines the entropy 
\begin{equation}\label{entropy def}
    S_w=-\frac{\partial F_w}{\partial T_w}\bigg\rvert_{X_w},
\end{equation}
and the dilaton chemical potential 
\begin{equation}\label{chemical potential0}
    \psi_w=-\frac{\partial F_w}{\partial X_w}\bigg\rvert_{T_w}.
\end{equation}

We also ensure thermodynamic stability of our black hole systems by verifying that their specific heat at constant dilaton charge is positive. This guarantees as well that the canonical ensemble and the saddle point approximation for the partition function are well defined.

In order to attain this aim we employ the definition for the specific heat at constant dilaton charge given by
\begin{equation}\label{specific heat}
    C_w= \frac{\partial E_w}{\partial T_w} \bigg\rvert_{X_w}.
\end{equation}

\subsubsection{Solution I}
Applying these definitions to our black hole solution I, we first use equations (\ref{tolman factor}), (\ref{reg action}), (\ref{free energy 1}) and the fact that
\begin{equation} \label{dilaton I}
X_w=x_0\left(r_{w}-\frac{c_1}{2}\right),   
\end{equation}
to calculate $F_w$ and obtain
\begin{equation}\label{energía libre}
    F_w=2x_0 \left(\frac{X_w}{x_0}-\frac{T}{T_w}-\frac{2\pi}{x_0}X_+ T_w\right).
\end{equation}

Based on (\ref{energía libre}) and (\ref{entropy def}) we compute the entropy of the black hole as follows\footnote{This expression for the entropy possesses a factor of 2 compared to (\ref{EntropyG}) since our action (\ref{I_reg onshell action}) also has this factor when compared to the action given in \cite{Grumiller}.}
\begin{equation}\label{entropía}
    S_w= S=-\frac{\partial F_w}{\partial T_w}\bigg\rvert_{X_w}=4\pi X_+=  4\pi x_0 \sqrt{\frac{c_1^2}{4}-c_2},
\end{equation}
where we have used the relation\footnote{This identity is obtained from equation (\ref{tolman factor}) by rewriting the Tolman factor in terms of $X_w$ and $X_+$.}
\begin{equation} \label{T and Tw with X}
\frac{T}{T_w}= \frac{1}{x_0}\sqrt{X_{w}^2-X_{+}^2}.
\end{equation}
We observe that the entropy of the black hole does not depend on the location of the wall $r_w$ but on the value $X_+$ of the dilaton at the horizon, just as in higher dimensions it depends on the area of the horizon. A universal form for the expression of the entropy is noted in (\ref{entropía}) when compared to other two-dimensional dilaton gravity models \cite{Grumiller,Davis,Nappi,Gegenberg,Grumiller4}.

As we deduced in (\ref{chemical potential0}), the chemical potential $\psi_w$ associated to the conserved charge (\ref{dilaton charge}) is 
\begin{equation}\label{potencial químico}
    \psi_w=-\frac{\partial F_w}{\partial X_w}\bigg\rvert_{T_w}=2 \left( \frac{T_w}{T}\frac{X_w}{x_0}-1\right),
\end{equation}
where relations (\ref{energía libre}) and (\ref{T and Tw with X}) were used. 

Following Brown and York \cite{Brown} we derive the quasi-local energy $E_w$ from the surface stress-energy-momentum tensor 
\begin{equation} \label{Brown stress tensor}
    T^{ij}:=- \frac{2}{\sqrt{\gamma}} \frac{\delta \Gamma}{\delta \gamma_{ij}},
\end{equation}
 by contracting $T^{ij}$ with $\xi_i\;\xi_j$, being $\xi_{i}=\sqrt{g_{tt}}\; \delta^\tau_{i}$ the Killing vector related to time translations. Varying the action (\ref{action2}) we encounter that
\begin{equation}
    T^{ij}=\frac{2x_0}{\gamma_{ij}}\left(\frac{X_w}{x_0}-\frac{T}{T_w}\right),
\end{equation}
accordingly
\begin{equation}\label{internal energy}
\begin{aligned}
    E_w=\xi_i\;\xi_j T^{ij}=2x_0\left(\frac{X_w}{x_0}-\frac{T}{T_w}\right) \geq 0,
\end{aligned}    
\end{equation}
where the restrictions (\ref{caso1i}) and (\ref{caso1ii}) have been used to assert that $E_w$ is positive or zero.

On the other hand, performing a Legendre transformation on (\ref{Legendre trans}) we obtain that the internal energy $E_w$ should obey 
\begin{equation} \label{legendre trans 2}
    E_w(S, X_w)=F_w(T_w, X_w)+ T_w \;S. 
\end{equation}
Hence, from equations (\ref{energía libre}) and (\ref{entropía}) we found that
\begin{equation} \label{internal energy 2}
    E_w(S,X_w)=2x_0\left(\frac{X_w}{x_0}-\frac{T}{T_w}\right),
\end{equation}
and we see that the internal energy deduced in this manner is in complete agreement with the result (\ref{internal energy}).

Using the identity (\ref{T and Tw with X}) and the expressions for entropy (\ref{entropía})  and dilaton chemical potential (\ref{potencial químico}) we verify the relation (\ref{first law0}) for the internal energy (\ref{internal energy}) by showing that it obeys the first law of black hole Thermodynamics. 

By taking  the differential of (\ref{internal energy}) we find that  
\begin{equation} \label{first law}
\begin{split}
    dE_w &= 2x_0 \left[ d\left(\frac{X_w}{x_0}\right)-d\left(\frac{T}{T_w}\right)\right]= 2x_0 \left[ d\left(\frac{X_w}{x_0}\right)-d\left(\frac{1}{x_0}\sqrt{X_{w}^2-X_{+}^2}\right)\right]\\
    &= 2 \left[ \left(1-\frac{X_w}{\sqrt{X_{w}^2-X_{+}^2}}\right)dX_w+\frac{X_+}{\sqrt{X_{w}^2-X_{+}^2}} dX_+\right]\\
    &= 2 \left[ \left(1-\frac{T_w}{T} \frac{X_w}{x_0}\right)dX_w+\frac{T}{\frac{1}{x_0}\sqrt{X_{w}^2-X_{+}^2}} \frac{dS}{2}\right]\\
    &= -\psi_w\;dX_w+T_w\;dS,
\end{split}    
\end{equation}
where in the third equality we can track back how the divergences in $\psi_w$ and $T_w$ cancel each other at $r_w=r_+$, that is to say, at $X_w=X_+$, verifying that $dE_w$ remains regular for all $r_w \geq r_+$, while in the
fourth equality we used $X_+=2\pi x_0 T$. Here the subscript $w$ indicates us that equation (\ref{first law}) remains valid no matter where the cavity wall is located along the $r$ coordinate. 

Relation (\ref{first law}) is one of the main results of this section and shows that our black hole configuration I possesses a consistent Thermodynamics.

It is important to note that under the extremality condition (\ref{extrem relation}) the entropy (\ref{entropía}) of this black hole vanish, recall that $X_+=x_0 \sqrt{\frac{c_1^2}{4}-c_2}$. Furthermore, it is easy to see that when using relation (\ref{extrem relation}), $\psi_w=E_w=T_w=0$. Hence, the first law is trivially fulfilled in the extremal case.

{\bf Black hole mass.}
Employing the ADM ($1+1$) decomposition, we compute the Hamiltonian for the Lorentzian  version of the action (\ref{action2}) and arrive at
    \begin{equation}\label{Hamiltonian}
      H=\int_{\Sigma_t} d r\left(N \mathcal{H}+N^r \mathcal{H}_r\right)+\left.\left(N \epsilon +N^r P_{rr}\right)\right|_B,
    \end{equation}
 here we foliate the spacetime in space-like hypersurfaces $\Sigma_t$ with boundary $B$, $N$ represents the lapse function and $N^r$ is the shift vector, $\mathcal{H}$ and  $\mathcal{H}_r$ are the Hamiltonian and momentum constraint respectively, $P_{rr}$ denotes the canonical momenta conjugate to the induced metric $g_{rr}$ in $\Sigma_t$. We verify that $\epsilon$ corresponds to the quasi-local energy $E_w$, defined in the second equality of equation (\ref{internal energy}).
 
 Evaluating the Hamiltonian (\ref{Hamiltonian}) with a solution of the field equations we obtain 
 \begin{equation}\label{on-shell H}
      H=\left.\left(N \epsilon +N^r P_{rr}\right)\right|_B.
    \end{equation}
As stated in \cite{Horowitz2}, this represents the total energy $M$ for spacetimes whose lapse function does not asymptotically approach unity.

Substituting our solution in equation (\ref{on-shell H}), noting that $P_{rr}=0$ for a static solution and $N=\sqrt{|g_{tt}|}$, we arrive at the total energy of the black hole
 \begin{equation}\label{M solI}
      M=N \epsilon \left. \right|_B= \lim_{r_w \rightarrow \infty} N\; E_w = x_0 \left(\frac{c_1^2}{4}-c_2 \right)= \frac{X_+^2}{x_0},
    \end{equation}
where we observe that the two constants of integration in the blackening function (\ref{blackening factor}) are involved in the definition of the black hole mass. As the last equality states, the mass is proportional to the squared dilaton evaluated at the event horizon.

From (\ref{M solI}) we easily see that the internal energy $E_w$ is asymptotically equal to the mass $M$, red-shifted by the Tolman factor:
 \begin{equation}
     \lim_{r_w \rightarrow \infty} E_w = \lim_{r_w \rightarrow \infty}  \frac{M}{r_w \sqrt{f(r_w)}}.
 \end{equation}

{\bf Specific heat.} By using the aforementioned definition for the specific heat (\ref{specific heat}) and the internal energy of this black hole solution (\ref{internal energy 2}) we arrive at
\begin{equation} \label{speheat I}
    C_w=\frac{2}{x_0}\;\frac{X_w^2-X_+^2}{T},
\end{equation}
where we have employed the expression for the Tolman factor $\frac{T}{T_w}=\frac{1}{x_0} \sqrt{X_w^2-X_+^2}$. As we see from (\ref{speheat I}) the specific heat for this solution is always positive or zero, given that $X_+ \leq X_w$ by design, yielding a stable configuration

Recalling that the solution I reproduces, as a particular case when $c_1=0$, the two-dimensional AdS black hole configuration in the group of the $a$-$b$ family of solutions with $b=1$, given in \cite{Katanaev, Grumiller}; and taking into account that the solution I accomplishes the Smarr-like formula
\begin{equation} \label{Smarr}
    M=\frac{TS}{2},
\end{equation}
we observe that the Thermodynamics of the black hole solution I and the one for the aforementioned $a$-$b$ solution are equivalent. For a given Hawking temperature (\ref{hawking temperature with surface g}), the two configurations present the same entropy (\ref{entropía}) and therefore the same energy (\ref{Smarr}).

\subsubsection{Solution II} \label{section 4.2.2}
In the case of the solution II, the counter-term action  (\ref{counter-term}) obtained by the Hamilton-Jacobi method is identically zero, since $\Lambda=0$, see equation (\ref{counterterm}). However, computing the on-shell action for this solution we find that
\begin{equation}
   \Gamma_{\text{reg}}= \Gamma=-x_0\frac{c_1}{T},
\end{equation}
is defined  by the geometry of the black hole configuration and is a constant value independent of the position of the regulatory boundary $r_{\text{reg}}$. Moreover, evaluating $\delta \Gamma$ for this solution, preserving the boundary conditions, we find a null variation. Given these properties, we are allowed to employ the semi-classical approximation (\ref{z with Gamma}) for the partition function.  In this case, the system  has a constant dilaton field $X=x_0$ and, thus, a trivial  dilatonic charge.

Making use of the same definitions as above, we deduce the Helmholtz free energy
\begin{equation}
    F_w(T_w,X_w)=-T_w \Gamma=x_0 c_1 \frac{T_w}{T},
\end{equation}
which we employ to compute the entropy
\begin{equation}
    S_w= S=-\frac{\partial F_w}{\partial T_w}\bigg\rvert_{X_w}= -x_0\frac{c_1}{T}=-4\pi  \frac{c_1}{\sqrt{c_1^2-4c_2}}x_0,
\end{equation}
which is a constant quantity proportional to the value of the constant dilaton $x_0$. With these two last results and making use of the Legendre transformation (\ref{legendre trans 2}) we deduce that the internal energy
\begin{equation}
    E_w=0,
\end{equation}
for this black hole solution. This is corroborated by the definition of Brown and York
\begin{equation}
    E_w= \xi_i\;\xi_j T^{ij}=-2 \; \xi_i\;\xi_j \pi ^{ij}=0,
\end{equation}
where in the second equality we make use of the relation (\ref{momenta as derivatives}) and the definition (\ref{Brown stress tensor}). In the third equality we employ the result
\begin{equation}
    \pi^{ij}=\gamma^{ij} n^{\mu} \nabla_{\mu} X,
\end{equation}
for a constant dilaton $X$.

Consequently, with the results for $E_w$ and $S$, we easily  realize that the Thermodynamics of this system is consistent in a trivial manner
\begin{equation}
    dE_w=T_wdS=0,
\end{equation}
as expected for two-dimensional models with constant dilaton.

Besides, by making use of the definition (\ref{specific heat}) we trivially obtain a null $C_w$, rendering a stable black hole solution.

From this last result and the corresponding outcome (\ref{speheat I}), we conclude that the specific heat at constant dilaton charge is either positive or zero for our solutions, independently of the location of the cavity wall $X_w$ employed in the canonical ensemble. In other words, there is no critical value $X_{\text{crit}}>X_+$ at which the specific heat becomes negative. This fact ensures that our black hole field configurations are thermodynamically stable without the need of a cavity wall.

\section{Our setup within dilatonic action frameworks} \label{sec5}

We would like to recall that the action (\ref{action_diagonal}) involving two scalar fields
\begin{equation} \label{action_u}
    S=\int \mathrm{d}^2 x \sqrt{-g} e^{\gamma_1 \phi_1+\gamma_2 \phi_2}\left[R+\beta_{1} \partial^\mu \phi_1 \partial_\mu \phi_1+\beta_{2} \partial^\mu \phi_2 \partial_\mu \phi_2-2 \Lambda\right], 
\end{equation}
upon substitution of the solution I, possesses a null kinetic term, allowing its expression in the dilatonic form \cite{Grumiller} 
\begin{equation} \label{action_g}
   I_d=\int \mathrm{d}^2 x \sqrt{-g}\left[X R-U(X) \partial^\mu X \partial_\mu X-2 V(X)\right],
\end{equation}
where $X\equiv e^{ \phi_1}$ and the potential functions of the dilaton field 
\begin{align} \label{potential}
&U(X)=0, & &V(X)=X\Lambda.   
\end{align}
From these relations, we realize that we are considering the Jackiw-Teitelboim theory \cite{Teitelboim, Jackiw, Cabrera}.

On the other hand, a solution-generating scheme for the field equations derived from the action (\ref{action_g})
\begin{equation} \label{eq g}
\begin{aligned}
U(X) \nabla_\mu X \nabla_\nu X-\frac{1}{2} g_{\mu \nu} U(X)(\nabla X)^2-g_{\mu \nu} V(X)+\nabla_\mu \nabla_\nu X-g_{\mu \nu} \nabla^2 X & =0, \\
R+\partial_X U(X)(\nabla X)^2+2 U(X) \nabla^2 X-2 \partial_X V(X) & =0,
\end{aligned}
\end{equation}
is presented, for instance in \cite{Grumiller,Grumiller2}, and is given by the following relations\footnote{Here the Euclidean signature is employed as in \cite{Grumiller2,Grumiller} for the sake of comparison.}:
\begin{equation}\label{sol1}
   X=X(r), \quad d s^2=\xi(r) d \tau^2+\frac{1}{\xi(r)} d r^2, 
\end{equation}
with
\begin{equation}\label{sol2}
\begin{aligned}
\partial_r X & =e^{-Q(X)}, \\
\xi(X) & =w(X) e^{Q(X)}\left(1-\frac{2 M}{w(X)}\right),
\end{aligned}
\end{equation}
given that 
\begin{equation}\label{soln3}
\begin{aligned}
& Q(X):=Q_0+\int^X dY\;U(Y), \\
& w(X):=w_0-2 \int^X dY\;V(Y) e^{Q(Y)} .
\end{aligned}
\end{equation}

Thus, by further considering the dilaton field  
\begin{equation}\label{dil1}
    X=x_0 \left(r-\frac{c_1}{2}\right),
\end{equation}
of the solution I, recalling that $x_0=e^{c_3+c_4}$, and by implementing the solution-generating scheme for solutions (\ref{sol1})-(\ref{soln3}) in order to obtain the metric function 
\begin{equation}\label{solnI}
    \xi(r)=\left(1-\frac{c_1}{r}+\frac{c_2}{r^{2}}\right)r^{2},
\end{equation}
where $c_1$ and $c_2$ are real constants of integration, we get our full {\it asymptotically $AdS$ black hole} solution with $\Lambda=-\frac{1}{l^2}$. Therefore, our field configuration 
solves the field equations (\ref{eq g}) derived from the JT action. As a consequence of the identification of these actions, the thermodynamic properties for this solution can be obtained following the method presented in \cite{Grumiller2,Grumiller}. 

We would like to notice that in \cite{Grumiller} it was shown that the metric function $\xi(X)$, derived in accordance to (\ref{sol2}), is parameterized by a {\it single} constant of integration, while in our solution, (\ref{solnI}) is determined by {\it two} integration constants. To the best of our knowledge, this solution has not been explicitly presented and studied elsewhere within the framework of the JT theory.

It is worth mentioning as well that in \cite{Grumiller2,Grumiller,Klosch} the authors report a similar {\it asymptotically flat black hole} solution with two constants of integration in the metric with a designed potential of the form 
    \begin{equation}\label{VX2}
        V(X)=-\frac{2 M}{X^2}+\frac{Q^2}{4 X^3},
    \end{equation}
where $M$ and $Q$ are constants of integration. 

The difference in the asymptotic properties of these black hole field configurations originates in the distinct potential functions (\ref{potential}) and $(\ref{VX2})$ of the dilaton field employed when constructing the corresponding solutions.

For the solution II, the metric function $\xi(r)$ also has the form (\ref{solnI}) and the scalar field configuration (\ref{scalars II}) leads to a constant dilaton field $X=x_0$, and therefore a trivial kinetic term, with no cosmological constant. 

One way to see that the action (\ref{action_g}) and the corresponding field equations lead to a different family of solutions from our solution II consists in substituting the latter conditions ($\nabla_{\mu}X=0=V(X)$) into  (\ref{eq g}) and comparing the properties of the resulting field configuration with those of our solution II. Thus, by doing this we obtain a configuration with $R=0$, while solution II corresponds to an $AdS$ black hole with constant and negative curvature $R=-\frac{2}{l^2}$, unveiling the novelty property of our solution.

{\bf Two-Dilaton Theories.} Finally, we would like to note that two dimensional dilaton gravity models of the form (\ref{action_u}) have been studied previously within the so-called Two-Dilaton Theories (TDT) classification \cite{Grumiller3}. In that work the following action is presented
\begin{equation}
\begin{aligned}
S_J= & \int_{M_2} d^2 x \sqrt{-g}\left[V_0^J(X, Y) R+V_1^J(X, Y) \nabla_\alpha X \nabla^\alpha X+V_2^J(X, Y) \nabla_\alpha Y \nabla^\alpha Y \right. \\
& \left. +V_3^J(X, Y) \nabla_\alpha X \nabla^\alpha Y +V_4^J(X, Y)+V_5^J(X, Y) f_m\left(S_n, \nabla_\alpha S_n, \ldots\right)\right],
\end{aligned}
\end{equation}
where $R$ is the curvature scalar, $X$ and $Y$ are the dilaton fields, the functions $V_i^J(X,Y)$ define the theory, $f_m$ represents functions of some matter fields $S_n$. By employing the following identifications
\begin{equation} \label{relations}
\begin{aligned}
    &V_0^J=XY, \quad V_1^J=\frac{Y}{X}\beta_{1}, \quad V_2^J=\frac{X}{Y}\beta_{2}, \quad V_3^J=0, \quad V_4^J=-2 \Lambda XY, \quad V_5^J=0, \\
    &\text{with} \quad X=e^{\phi_1}, \quad Y=e^{\phi_2}.
\end{aligned}    
\end{equation}
we realize that the system (\ref{action_u}) corresponds to a particular case of the models studied in the above reference.

\subsection{Constant dilaton vacua in our setup and their stability}

     In order to study the possibility that our black hole solutions decay into field configurations with lower free energy within our setup, we look for constant dilaton vacua (CDV) \cite{Grumiller} with the same boundary conditions  in the canonical ensemble for our two solutions.

     CDV are solutions that accomplish the field equations (\ref{eq g}) with a constant boundary condition for the dilaton field $X=X_0$ that renders $V(X)=0$, a metric function (\ref{sol2}) given by 
    \begin{equation} \label{metric cdv}
        \xi=c+ar-\frac{1}{2}\lambda r^2,
    \end{equation}
where $a$, $c$ and $\lambda$ are constants, and the following condition for the curvature scalar
\begin{equation} \label{scalar cdv}
R^{CDV}=-\partial_r^2 \xi=\lambda=2 \partial_XV(X)\big\rvert_{X_0} ,
\end{equation}
which determines $\lambda$.
This boundary condition is valid for both the CDV and our black hole solutions in the canonical ensemble.

On the other hand, the canonical ensemble is determined by a cavity wall located out of the event horizon $X_w>X_+$, defining the same boundary condition for the dilaton, therefore $X_0=X_w$, and hence $X_0>X_+$.

Consequently, we further need to express the dilaton field evaluated at the outer horizon $X_{+}$ of a given black hole solution and compare it to $X_0$ in order to elucidate whether the canonical ensemble and the corresponding free energy of the CDV are well-defined. If this is the case, we finally need to compare the free energies of both field configurations in order to establish which one is thermodynamically favored, i. e. whether tunneling from a black hole solution into a CDV or viceversa is a favorable process.

For the solution I of our model, we see from the second relation in (\ref{potential}) that the restriction $V(X)=0$ necessarily implies $X_0=0$. Given that the horizon is positive by definition $X_{+} > 0$, we see that for this case $X_{+} > X_0$, meaning that the obtained CDV solution does not possess a meaningful free energy since the cavity wall would be inside the event horizon. 

  Although, it is impossible to reduce the action (\ref{action_u}) into (\ref{action_g}) for the configuration II, we look for an analogous solution to the CDV by employing constant scalar fields $\phi_1$ and $\phi_2$, arriving to a null scalar curvature $R=0$, a result that differs from the negative and constant scalar curvature of our black hole configuration. This implies that the boundary conditions for the analogous CDV differ from the boundary conditions of solutions II in the same canonical ensemble, avoiding the comparison of their free energies.

 Thus, our model does not exhibit appropriate CDV with a well-defined free energy that can be compared to the free energy of our black hole solution and investigate the possibility of tunneling from one configuration into another.

\section{Boundary effective theory} \label{sec6_0}

In this section, we study the effective theory living at the boundary of our $AdS_2$ black hole solution I. With this purpose, we stay in Euclidean signature and consider the change of coordinates $r \rightarrow 1/z$ and  $t \rightarrow t$, such that we express the metric (\ref{blackhole I}) and the dilaton field (\ref{dilaton I}) as follows
\begin{align}
    &ds^2= \frac{l^2}{z^2} \left(f(z) dt^2 + \frac{dz^2}{f(z)}\right), \quad f(z)=1-c_1 z+c_2 z^2, \label{Poincaré metric}\\ 
    &X(z)= x_0 \left( \frac{1}{z} - \frac{c_1}{2}\right). \label{Poincaré dilaton}
\end{align}    

Although the boundary is now located at $z=0$, we consider to cut-off the spacetime along a trajectory given by the parametrized curve $(t(u),z(u))$, where the parameter $u$ is usually interpreted as the time in the boundary theory.  From (\ref{Poincaré metric}) we have that the induced metric $g_{uu}$ on this curve is given by
\begin{equation}
    ds^2=\frac{l^2}{z^2} \left( f(z) \dot{t}^2+ \frac{\dot{z}^2}{f(z)} \right) du^2 = g_{uu} du^2,
\end{equation}
where a dot denotes a derivative with respect to the parameter $u$. 

If we consider the spacetime cut-off near the boundary, where $z \rightarrow 0$ and therefore
\begin{equation}\label{met boun}
    g_{uu} \rightarrow g_{uu}|_{bdy}=\frac{l^2}{z^2}\left(\dot{t}^2+ \dot{z}^2 \right), 
\end{equation}
we demand that, following \cite{Stanford}, the induce metric at this new boundary trajectory 
\begin{equation}\label{cond met bound}
    g_{uu}|_{bdy}=\frac{1}{\epsilon^2}, 
\end{equation}
where $\epsilon$ is a small parameter. This condition leads 
to the relation
\begin{equation} \label{boundary z}
    z=l\epsilon \; \dot{t} + O\left(\epsilon^3 \right).
\end{equation}

Evaluating our action 
\begin{equation}\label{action4}
\begin{aligned}
        \Gamma &= \int _{\mathcal{M}} d^{2} x \sqrt{g} \; X \left ( R + \sum_{b} \beta_{b} (\partial^{\mu} \phi_b)(\partial_{\mu} \phi_b) - 2 \Lambda \right )
        +2\int_{\partial \mathcal{M}} d x \sqrt{\gamma}\; X\; K \\
        &- C\int_{\partial\mathcal{M}} dx \sqrt{\gamma}\; X,
\end{aligned}        
\end{equation}
with solution I, the bulk term vanishes identically, only the surface terms contribute to the on-shell action.

Computing the trace of the extrinsic curvature $K$ at a parametrized curve $(t(u),z(u))$ results in
\begin{equation}\label{extrinsic curv}
\begin{split}
    K= &\frac{\sqrt{f}}{l \sqrt{\dot{t}^2+\dot{z}^2 f^2}\left(\dot{t}^2 f^2+\dot{z}^2 \right)}* \\
    &\biggl\{ f^2 \dot{t} \left(\dot{t}^2+\dot{z}^2+z \ddot{z} \right) -z \dot{z} \ddot{t} -z \dot{t}\left[\frac{ff'}{2} \left(\dot{t}^2-\dot{z}^2\right) + \left(f^2-1\right) \dot{z} \frac{\dot{t}\ddot{t}+\dot{z}\ddot{z}f^2+\dot{z}^3ff'}{\dot{t}^2+\dot{z}^2 f^2}\right]   \biggr\},
\end{split}
\end{equation}
here, $f=f(z)$ is the blackening factor in (\ref{Poincaré metric}) and the prime denotes a derivative with respect to  the variable $z$.

Using the boundary relation (\ref{boundary z}) in (\ref{extrinsic curv}) we obtain the trace of the extrinsic curvature at the boundary curve
\begin{equation}\label{boundary k}
    K|_{bdy}=\frac{1}{l} \biggl\{1+\left[Sch(t(u),u)+M\;\frac{\dot{t}^2}{2x_0}\right]l^2 \epsilon^2 + O\left(\epsilon^3 \right) \biggr\},
\end{equation}
where the $Sch(t(u),u)$ term is the $SL(2,R)$ invariant Schwarzian derivative \cite{Kitaev, Stanford3}
\begin{equation}
Sch(t(u),u)=\frac{2\dot{t}\;\dddot{t}-3\ddot{t}^2}{2\dot{t}^2},
\end{equation}
which is supplemented by a new finite temperature term parameterized by the mass $M$ of the black hole solution I
\begin{equation}\label{tilde M}
    M=x_0\left(\frac{c_1^2}{4}-c_2\right).
\end{equation}
It is worth noting that, in the construction of (\ref{boundary k}), the $\epsilon^2$ term collects an additional black hole mass contribution determined by the corresponding constants of integration, as shown in (\ref{tilde M}). 
Consequently, these integration constants will characterize the effective boundary action and its corresponding solutions.
The aforementioned black hole mass term apparently breaks the $SL(2,R)$ invariance of the boundary trace of the extrinsic curvature (\ref{boundary k}). However, it will be shown that this relevant symmetry becomes explicit under a suitable time reparameterization that involves the black hole mass, as it will be discussed below.

On the other hand, in the limit when the black hole mass vanishes, we recover the Schwarzian derivative of pure $AdS_2$ gravity \cite{Almheiri,Stanford}, as expected. More importantly, the same boundary Schwarzian derivative arises for extremal $AdS_2$ black hole configurations that satisfy the relation (\ref{extrem relation}), a novel situation that takes place thanks to the existence of the two integration constants, $c_1$ and $c_2$, in the blackening factor of our solution.

Moreover, considering  (\ref{boundary z}) into (\ref{Poincaré dilaton}) we obtain the boundary condition for the dilaton field 
\begin{equation} \label{X bound}
    X|_{bdy}=\frac{x_0}{l\dot{t}\epsilon}. 
\end{equation}

Finally, we compute the effective action by substituting into the surface terms of the action (\ref{action4}) the boundary value (\ref{boundary k}) and the boundary conditions (\ref{cond met bound}) and (\ref{X bound})
\begin{equation}\label{action5}
\begin{aligned}
        \Gamma &= 2\int_{\partial \mathcal{M}} \frac{du}{\epsilon}\; X|_{bdy}\; K|_{bdy} - \frac{2}{l}\int_{\partial\mathcal{M}} \frac{du}{\epsilon}\; X|_{bdy}\\
        &= 2x_0\int_{\partial\mathcal{M}} \frac{du}{\dot{t}} \left[Sch(t(u),u)+ M\;\frac{\dot{t}^2}{2x_0}\right]\\
        &= 2x_0\int_{\partial\mathcal{M}} d\tilde{u} \left[Sch(t(u),u)+ M\;\frac{\dot{t}^2}{2x_0}\right],
\end{aligned}        
\end{equation}
here, we have substituted the value $C=2/l$ for the solution I counter-term, computed in section \ref{sec 4.2}. Moreover, we observe that the divergent item in the Gibbons-Hawking-York term is removed by the counter-term. On the other hand, we can always remove the overall factor $1/\dot{t}$ in the second line by considering a new boundary time coordinate with $du=d\tilde{u}\;\dot{t}$, as we do in the last line.

If we consider the function
\begin{equation} \label{comp t}
    \tau(u)= tan\left(\sqrt{\frac{M}{4x_0}}t(u)\right),
\end{equation}
we are allowed to implement the composition rule for the Schwarzian of a composite function $f\circ g(u)$
\begin{equation}
    S(f\circ g,u)=S(g(u),u)+\dot{g}^2 S(f(g),g),
\end{equation}
to obtain that the effective action adopts the form
\begin{equation} \label{transf Sch}
    \begin{aligned}
        \Gamma&= 2x_0\int_{\partial\mathcal{M}} du \left[Sch(t(u),u)+M\;\frac{\dot{t}^2}{2x_0}\right]=2x_0\int_{\partial\mathcal{M}}du \left[Sch(t(u),u)+\dot{t}^2\;Sch(\tau(t),t)\right] \\
        &=2x_0\int_{\partial\mathcal{M}} du \left[Sch(\tau(u),u)\right].
    \end{aligned}
\end{equation}
Thus, by varying with respect to $\tau(u)$ the action in the last line, we deduce the equation of motion for our effective action 
\begin{equation} \label{Sch eq mot}
    \frac{\dot{Sch}(\tau(u),u)}{\dot{\tau}}=0.
\end{equation}
From (\ref{comp t}) and (\ref{transf Sch}), we realize that if $t(u)$ is a linear function of $u$ we have a non-constant $\tau(u)$ with a constant $Sch(\tau,u)$, therefore, we have a family of solutions to equation (\ref{Sch eq mot}). Moreover, recalling from section \ref{sec 4.1}, our Euclidean time $t(u)$ is periodic with period $1/T=2 \pi \sqrt{\frac{x_0}{M}}$. Hence, the family of solutions for (\ref{Sch eq mot}) are given by (\ref{comp t}) with
\begin{equation} \label{t sol}
    t(u)=2 \pi \sqrt{\frac{x_0}{M}}\;\frac{u}{\tilde{\beta}}=\frac{u}{T\tilde{\beta}},
\end{equation}
where the periodicity of $t(u)$ demands the periodicity of the boundary time $u$, with period $\tilde{\beta}$, this leads us to the boundary temperature $\tilde{T}=1/\tilde{\beta}$.  

Evaluating the action (\ref{transf Sch}) with the solutions (\ref{t sol}) we obtain the on-shell action
\begin{equation} \label{onshell bound theo}
    \begin{aligned}
    &\Gamma=2\,\tilde{C}\;\frac{\pi^2}{\tilde{\beta}}, \quad &\tilde{C}=2\,x_0. 
     \end{aligned}
\end{equation}

In order to obtain the Thermodynamics for the boundary theory, we apply the definitions in section \ref{sec 4.2.1}. First, we compute the Hemholtz free energy  
\begin{equation}
     F=-\tilde{T}\;\Gamma=-\tilde{C}\;2\pi^2 \tilde{T^2}.
\end{equation}
With this result we derive the entropy $S$, energy $E$ and specific heat $C$ as follows
\begin{equation}
\begin{aligned}
    &S=-\frac{\partial F}{\partial \tilde{T}}=\tilde{C}\; 4\pi^2 \tilde{T}, \quad &E=F+\tilde{T}S=\tilde{C} \;2\pi^2 \tilde{T}^2=Sch(\tau,u), \\
    &C=\frac{\partial E}{\partial\tilde{T}}=\tilde{C}\;4\pi^2\tilde{T}.
\end{aligned}    
\end{equation}

We would like to note here the following, even though our solutions (\ref{t sol}) are determined by the Hawking temperature $T$ and therefore by the two integration constants $c_1$ and $c_2$, the resulting on-shell action (\ref{onshell bound theo}) is not different from the one for an $AdS_2$ spacetime without a black hole causal structure, leading to the same thermodynamical properties for the boundary theory.

 Given that, under the extremality condition (\ref{extrem relation}) we have a null black hole mass $M$ and Hawking temperature $T$, the effective action (\ref{transf Sch}) is only given by the $Sch(t(u),u)$ and the Euclidean time $t(u)$, with $0\leq t(u)\leq 1/T$, is no longer periodic. Thus, in order to obtain the Thermodynamics for the boundary theory, we follow \cite{Stanford} and change to the Euclidean Rindler time, setting $t(u)=tan\frac{\tau(u)}{2}$, with $\tau$ periodic with period $2\pi$. Then we apply the composition law for the Schwarzian and obtain
\begin{equation} \label{transf Sch2}
    \begin{aligned}
        \Gamma&= 2x_0\int_{\partial\mathcal{M}} du \left[Sch(t(u),u)\right]=2x_0\int_{\partial\mathcal{M}} du \left[Sch(\tau(u),u)+\frac{\dot{\tau}^2}{2}\right].
    \end{aligned}
\end{equation}
Similarly as above, this tells us that the solutions to the equation of motion for this action are given by 
\begin{equation} \label{t sol ex}
    \tau(u)=\frac{2\pi}{\tilde{\beta}}\;u.
\end{equation}
Substituting these solutions into (\ref{transf Sch2}), we obtain the same on-shell action (\ref{onshell bound theo}) and therefore the same Thermodynamics for the extremal case.

A final remark is warranted regarding the holographic description of the boundary theory in solution II. As shown in Sec. \ref{section 4.2.2}, the dilaton is constant in this case, leading to a vanishing spacetime energy. The dual theory is a trivial one-dimensional conformal field theory ($CFT_1$) with no dynamical degrees of freedom \cite{Stanford}. For further discussion, see \cite{Grumiller5}, where the authors presented a comprehensive holographic analysis of a generic two-dimensional dilaton–gravity model coupled to a Maxwell field, which admits $AdS_2$ solutions with a constant dilaton field. By computing the canonical charges and the quantum gravity partition function, they demonstrate the triviality of $AdS_2$ holography under constant dilaton boundary conditions.

\section{Conclusions} \label{sec6}

In this paper we have presented two novel analytic $AdS$ black hole solutions with two constants of integration  in the blackening factor, for two-dimensional dilaton gravity models whose action is described by equation (\ref{action_diagonal}).

We have demonstrated in solution I that the scalar field configuration supporting the metric (\ref{blackhole I}), within a theory with a non-vanishing cosmological constant, must consist of only one non-trivial scalar field.

In the particular case when $c_1=0$, our solution I coincides with the two-dimensional AdS black hole configuration presented in \cite{Katanaev,Grumiller} as a part of the $a$-$b$ family of black hole solutions, when his parameter $b=1$.

For the presented solutions, we have extended the spacetime toward the interior of the black hole employing appropriate Kruskal-like coordinates, a construction developed in section \ref{sec3}, in order to prove the black hole nature of our black hole configurations. We have found a resemblance with the causal structure of the RN black hole, revealing the event and apparent horizon character of $r_+$ and $r_-$, respectively; we have illustrated this spacetime structure, in a compact form, with a Penrose diagram.

Under the extremality condition (\ref{extrem relation}) and making use of appropriate change of variables (\ref{our change of coord}), we were able to show that our solutions present an $AdS_2$ geometry outside the black hole, and not only at the near horizon region, as is the case for the extremal black hole configurations in higher dimensions, for example, the $4D$ extremal RN black hole.

We have deduced a formula for the Hawking temperature that can be expressed in terms of the surface gravity $\kappa$ for our solutions. In particular, this quantity turns out to be null in the case of the extreme configuration.

Finally, we have developed consistent Thermodynamics for the black holes presented in this work. For this purpose, we have employed the approximation (\ref{approximation z}) for the partition function $\mathcal{Z}$ in the canonical ensemble. In order to use this approximation, we have constructed a renormalized action $\Gamma$ (\ref{action2}) with the Hamilton-Jacobi method explained in detail in section \ref{sec 4.2}. The essential part of this method is the construction of a boundary counter-term (\ref{counter-term}) that removes the divergences in the on-shell action  and leaves a renormalized action that is certainly extremized by the solutions. In this procedure, we have defined the regulated on-shell action $\Gamma_{\text{reg}}$ (\ref{reg action}), with finite boundary $r_{\text{reg}}$, as a part of the limiting procedure for evaluating the on-shell action $\Gamma$. Besides, we have employed $\Gamma_{\text{reg}}$ as the argument of the exponential in the approximation (\ref{z with Gamma}) for the partition function. 

Following the approach of York \cite{York2}, we have think of the black hole as to be in a cavity or a box whose frontier is placed at the wall $r_{\text{w}}=r_{\text{reg}}$ that is in equilibrium with a thermal reservoir. In this manner, we have introduced the partition function in the canonical ensemble for our black hole configurations. This ensemble is defined by the local temperature $T_w$, which we have related to the Hawking temperature $T$ employing a Tolman factor in equation (\ref{tolman factor}), and a dilaton charge $D_w$ (\ref{dilaton charge}), introduced in section \ref{sec 4.2.1}. We have computed the Helmholtz free energy $F_w$ in order to calculate the entropy $S_w$ and the dilaton chemical potential $\psi_w$. 

As we noted above, the entropy for the black hole in solution I
\begin{equation}\label{entropy2}
    S_w= S= 4\pi X_+,
\end{equation}
does not depend on $r_w$ but on the value of the horizon $r_+$; this is analogous to the case in higher dimensions where the entropy depends on the area of the event horizon. We have found that the form of (\ref{entropy2}) is in accordance with distinct dilaton gravity models, for example \cite{Grumiller,Davis,Nappi}. Following Brown and York \cite{Brown}, we have derived, from the quasi-local stress tensor (\ref{Brown stress tensor}), the proper energy density given by (\ref{internal energy}). This internal energy is also encountered in the Legendre transformation (\ref{legendre trans 2}). Finally, we have verified that the first law of black hole Thermodynamics is fulfilled for the system enclosed by the cavity wall $r_w$, remaining regular for all $r_w\geq r_+$. This fact provides physical consistency to our black hole configuration I, rendering a viable model from the thermodynamic viewpoint. 

We have found that there is no counter-term for the black hole solution II, with cosmological constant $\Lambda=0$, in accordance with equation (\ref{counterterm}). Nonetheless, we have encountered the on-shell action for this solution to be adequate for the semiclassical approximation method with which we deduce a consistent Thermodynamics. In this case, the constant dilaton leads to trivial Thermodynamics.

We have calculated the mass $M$ of the black hole solutions, defined by the solution-valued Hamiltonian with a lapse function different from unity. For the black hole solution I, the mass is proportional to the squared dilaton evaluated at the event horizon $X_+^2$; particularly, $M$ depends on the two constants of integration $c_1$ and $c_2$ of the blackening function $f(r)$. 

We have computed the specific heat at constant dilaton charge, and we have found that $C_w$ is a positive or null quantity for our solutions, independently of the location of the cavity wall $X_w$. This fact ensures the thermodynamic stability of our field configurations without requiring a finite cavity.

Based on the Smarr-like formula (\ref{Smarr}) we found that solution I is thermodynamically equivalent to the $AdS_2$ black hole configuration in the $a$-$b$ family of solutions with $b=1$, given in \cite{Katanaev, Grumiller}. 
Requiring equality for the Hawking temperature $T$ in both configurations yields the same entropy $S$ and energy $M$.

We have performed a holographic study of the effective boundary theory of our $AdS_2$ black hole solution I. We have verified that the counter-term deduced for this bulk solution eliminates exactly the divergent term arising in the boundary action. This effective action is defined by a Schwarzian derivative plus a term involving the mass $M$ of the bulk theory. This fact leads us to have solutions $t(u)$ determined by the Hawking temperature $T$, and therefore by the two integration constants $c_1$ and $c_2$ of the blackening factor. Nevertheless, the boundary on-shell action for this solutions remains the same as the one for an $AdS_2$ spacetime without a black hole causal structure, producing in this manner the same Thermodynamics of a theory living at the boundary of a zero-temperature $AdS_2$ spacetime.

We would like to point out that it would be interesting to explore whether is it possible to generalize the results of this work to higher dimensions employing multiple scalar fields and considering anisotropy between the time and the spatial sector as in Lifshitz spacetimes. 

Finally, we note that while preparing this article, a parallel work was published \cite{Mao}, where similar results were obtained (in particular the metric of our solutions) within the two-dimensional dilaton gravity with a different scalar field setup (with potentials $U$ and $V$ that differ from ours). The effective scalar-tensor theory emerges as the $D \rightarrow 2$ limit of Einstein gravity with cosmological constant upon a Kaluza-Klein dimensional reduction.

\begin{acknowledgments}
All the authors are grateful to Manuel de la Cruz López, Jhony A. Herrera-Mendoza, Daniel F. Higuita-Borja, Julio A. Méndez-Zavaleta, Ulises Nucamendi, G. F. Torres del Castillo, Mehrab Momennia and Olivier Sarbach for fruitful and illuminating discussions. 
UNC acknowledges support from CONAHCYT through a PhD Grant No.814574, AHA has benefited from  CONAHCYT grants No. A1-S-38041 and No. CF-MG-2558591, and VIEP-BUAP. Finally, AHA and CRR thank the Sistema Nacional de Investigadoras e Investigadores (SNII) for support.
\end{acknowledgments}

\bibliography{main}
\end{document}